%% file: paper.tex
\documentclass[a4paper,UKenglish,cleveref, autoref, thm-restate, authorcolumns, nolineno]{gd-lipics-v2}

\usepackage{amsmath}
\usepackage{amssymb}
\usepackage{amsmath}
\usepackage{amssymb}
\usepackage{booktabs}
\usepackage[table]{xcolor}
\usepackage{colortbl}
\usepackage{lscape}
\usepackage{graphicx}
\usepackage{caption}
\usepackage{subcaption}
\usepackage{float}
\usepackage{pdfpages}
\usepackage{accsupp}
\usepackage{pdfcomment}

\hideLIPIcs

\title{Show Me Your Best Side: Characteristics of User-Preferred Perspectives for 3D Graph Drawings}

\author{Lucas Joos}{University of Konstanz, Germany}{lucas.joos@uni-konstanz.de}{https://orcid.org/0000-0001-7049-5203}{}

\author{Gavin J. Mooney}{Monash University, Australia}{gavin.mooney@monash.edu}{https://orcid.org/0009-0001-6208-4268}{}

\author{Maximilian T. Fischer}{University of Konstanz, Germany}{max.fischer@uni-konstanz.de}{https://orcid.org/0000-0001-8076-1376}{}

\author{Daniel A. Keim}{University of Konstanz, Germany}{keim@uni-konstanz.de}{https://orcid.org/0000-0001-7966-9740}{}

\author{Falk Schreiber}{University of Konstanz, Germany \and Monash University, Australia}{falk.schreiber@uni-konstanz.de}{https://orcid.org/0000-0002-9307-3254}{}

\author{Helen C. Purchase}{Monash University, Australia}{helen.purchase@monash.edu}{https://orcid.org/0000-0001-6994-4446}{}

\author{Karsten Klein}{University of Konstanz, Germany}{karsten.klein@uni-konstanz.de}{https://orcid.org/0000-0002-8345-5806}{}

\authorrunning{L. Joos et al.}

\Copyright{Lucas Joos, Gavin J. Mooney, Maximilian T. Fischer, Daniel A. Keim, Falk Schreiber, Helen C. Purchase, and Karsten Klein} 



\category{Track 2} 

\funding{The authors gratefully acknowledge financial support by the Federal Ministry for Economic Affairs and Climate Action (BMWK, grant No. 03EI1048D) and the Deutsche Forschungsgemeinschaft (DFG) – Project-ID 251654672 – TRR 161.}

\EventEditors{Vida Dujmovi\'c and Fabrizio Montecchiani}
\EventNoEds{2}
\EventLongTitle{33rd International Symposium on Graph Drawing and Network Visualization (GD 2025)}
\EventShortTitle{GD 2025}
\EventAcronym{GD}
\EventYear{2025}
\EventDate{September 24--26, 2025}
\EventLocation{Norrk\"{o}ping, Sweden}
\EventLogo{}
\SeriesVolume{357}
\ArticleNo{35}

\newcommand{\GLayout}[1]{\texttt{L-#1}}
\newcommand{\GSize}[1]{\texttt{S-#1}}

\newcommand{\MeasureNNO}{\pdftooltip{\textsf{NNO}}{Node-Node Overlap Count}}
\newcommand{\MeasureENO}{\pdftooltip{\textsf{ENO}}{Edge-Node Overlap Count}}
\newcommand{\MeasureNEO}{\pdftooltip{\textsf{NEO}}{Node-Edge Overlap Count}}
\newcommand{\MeasureNNOA}{\pdftooltip{\textsf{NNOA}}{Node-Node Overlap Area}}
\newcommand{\MeasureENOA}{\pdftooltip{\textsf{ENOA}}{Edge-Node Overlap Area}}
\newcommand{\MeasureNEOA}{\pdftooltip{\textsf{NEOA}}{Node-Edge Overlap Area}}
\newcommand{\MeasureCR}{\pdftooltip{\textsf{CR}}{Edge Crossings}}
\newcommand{\MeasureST}{\pdftooltip{\textsf{ST}}{Stress}}
\newcommand{\MeasureCAR}{\pdftooltip{\textsf{CAR}}{Crossing Angular Resolution}}
\newcommand{\MeasureAR}{\pdftooltip{\textsf{AR}}{Bounding Box Area}}
\newcommand{\MeasureASP}{\pdftooltip{\textsf{ASP}}{Aspect Ratio}}
\newcommand{\MeasureCON}{\pdftooltip{\textsf{CON}}{Node Concentration}}
\newcommand{\MeasureNO}{\pdftooltip{\textsf{NO}}{Node Orthogonality}}
\newcommand{\MeasureGR}{\pdftooltip{\textsf{GR}}{Gabriel Ratio}}
\newcommand{\MeasureANGR}{\pdftooltip{\textsf{ANGR}}{Angular Resolution}}
\newcommand{\MeasureEO}{\pdftooltip{\textsf{EO}}{Edge Orthogonality}}
\newcommand{\MeasureELD}{\pdftooltip{\textsf{ELD}}{Edge Length Deviation}}
\newcommand{\MeasureESR}{\pdftooltip{\textsf{ESR}}{Edge Reflective Symmetry}}
\newcommand{\MeasureESO}{\pdftooltip{\textsf{ESO}}{Edge Rotational Symmetry}}
\newcommand{\MeasureEST}{\pdftooltip{\textsf{EST}}{Edge Translational Symmetry}}
\newcommand{\MeasureISO}{\pdftooltip{\textsf{ISO}}{Isometric Viewpoint Deviation}}
\newcommand{\MeasureCLR}{\pdftooltip{\textsf{C-LR}}{Combination based on Logistic Regression}}
\newcommand{\MeasureCSQP}{\pdftooltip{\textsf{C-SQP}}{Combination based on Sequential Quadratic Programming}}

\begin{document}

\maketitle

\begin{abstract}
The visual analysis of graphs in 3D has become increasingly popular, accelerated by the rise of immersive technology, such as augmented and virtual reality.
Unlike 2D drawings, 3D graph layouts are highly viewpoint-dependent, making perspective selection critical for revealing structural and relational patterns.
Despite its importance, there is limited empirical evidence guiding what constitutes an effective or preferred viewpoint from the user’s perspective.
In this paper, we present a systematic investigation into user-preferred viewpoints in 3D graph visualisations.
We conducted a controlled study with 23 participants in a virtual reality environment, where users selected their most and least preferred viewpoints for 36 different graphs varying in size and layout.
From this data, enriched by qualitative feedback, we distil common strategies underlying viewpoint choice.
We further analyse the alignment of user preferences with classical 2D aesthetic criteria (e.g., \textit{Crossings}), 3D-specific measures (e.g., \textit{Node-Node Occlusion}), and introduce a novel measure capturing the perceivability of a graph’s principal axes (\textit{Isometric Viewpoint Deviation}).
Our data-driven analysis indicates that \textit{Stress}, \textit{Crossings}, \textit{Gabriel Ratio}, \textit{Edge-Node Overlap}, and \textit{Isometric Viewpoint Deviation} are key indicators of viewpoint preference.
Beyond our findings, we contribute a publicly available dataset consisting of the graphs and computed aesthetic measures, supporting further research and the development of viewpoint evaluation measures for 3D graph drawing.
\end{abstract}

\section{Introduction}
\label{sec:introduction}
Beyond traditional graph drawing in two-dimensional space, the rise of immersive technology has enabled numerous applications and studies where graphs are arranged and viewed in three-dimensional space using devices that support stereoscopic 3D (S3D) vision~\cite{joos2025visual}.
Virtual (VR) and augmented reality (AR) devices are increasingly employed for this purpose, due to their growing availability, quality, and affordability, across fields such as biology~\cite{kuznetsov2021the}, neuroscience~\cite{pirch2021vrnetzer}, and the social sciences~\cite{drogemuller2017vrige}.
Although rare approaches visualise graphs as matrices in 3D space~\cite{pan2024extending}, the common method remains node-link diagrams~\cite{joos2022visual} where node positions are typically derived from layout algorithms or reflect semantic 3D positions (such as brain regions~\cite{jaeger2019challenges}).
Under certain conditions, S3D environments have been shown to enhance graph-related tasks, such as community detection~\cite{greffard2014three} and path tracing~\cite{mcguffin2024path}.

However, due to the three-dimensional representation and the impact of depth and occlusion, viewpoints might either facilitate or prohibit the identification of important aspects and structures, thereby supporting or hindering the analysis process.
Therefore, 3D visualisations are often viewed from several viewpoints for interactive graph analysis, which is mostly done by the analyst in an ad-hoc fashion.
Investigating the impact of different viewpoints and their potential relation to layout characteristics and quality measures consequently deserves increased attention.
One pathway to derive insight on this impact is to investigate the variability of established 2D quality measures for a sample of viewpoints without considering human perception at all.
However, these measures might not be a good fit for the quality of perceived structures in a 3D environment.
A complementary approach is to investigate the characteristics of viewpoints chosen by humans. 

It is well-known that preference for a visualisation does not necessarily strongly correlate with better task performance or lower cognitive load~\cite{Huang09}.
Still, preferences that are not solely based on aesthetic appearance, but motivated by a context in graph analysis, can guide the evaluation of quality measures and facilitate the search for new ones.
Even if humans would sometimes prefer less advantageous viewpoints, it is an important goal to avoid viewpoints that are reasoned to be subpar, in particular if objective guidance for improvement can be given.
In interactive analysis, the choice of what to explore next can be influenced, or even misled, by the previous choice, especially if that earlier choice did not provide enough information to make reasonable decisions going forward.
Hence, optimised initial viewpoints can be especially valuable.
Further, identifying characteristics that lead to specific preferences allows us to gain further insight into human graph analysis behaviour.
In the literature, measures intended to capture the quality of a graph drawing are described as either \textit{aesthetic} or \textit{readability} measures, terms whose meanings partly overlap but do not always coincide.
As our study centres on measures grounded in viewers' preferences, we adopt the term \textit{aesthetic} throughout this paper.
Our investigation aims to better understand human preference for 3D viewpoint selection, investigating the following research question:
\vspace{2mm}\\
\noindent
\textbf{RQ:} \textit{What are users' preferences regarding viewpoints in 3D graph drawings, and to what extent can individual aesthetic measures (e.g., edge crossings, stress) or combinations of these characterise preferred or unfavoured perspectives?}
\vspace{2mm}

\noindent
Thereby, we make the following contributions:
\vspace{1mm}
\begin{itemize}
    \setlength\itemsep{1mm}
    \item A VR \textbf{user experiment} ($n=23$) collecting a set of user-preferred and -disregarded perspectives for a broad set of 36 3D graph drawings along with qualitative user feedback.
    \item A detailed analysis of the study results, consisting of a \textbf{qualitative and quantitative analysis} evaluating the expressiveness of individual aesthetic measures (21 in total, including one introduced in this work) and \textbf{combinations} of these.
    \item We publish our \textbf{study data} along with all results and analyses as \textbf{open-access}, providing a valuable dataset for future research on this topic: \href{https://doi.org/10.18419/DARUS-5108}{{\color{blue} 10.18419/DARUS-5108}}~\cite{joos2025Darus}.
\end{itemize}

\section{Related Work}
\label{sec:related-work}

There is some evidence that people, when presented with a static graph drawing, tend to prefer those with lower values for quality measures such as edge crossings or stress~\cite{ChimaniPoster}. However, in interactive settings, they do not necessarily optimise for the lowest values~\cite{marner2014gion}.
With the additional degree of freedom in 3D visualisations, this tendency may be even more pronounced due to both the limited applicability of classical quality measures to viewpoint evaluation and the difficulty humans face in judging layout quality for analytical purposes.
Over the years, many quality measures have been proposed~\cite{Bennett07,Dib24}, some of which are better supported by empirical studies than others.
A relevant question is to what extent viewers can actually perceive and distinguish different values of such measures, i.e., whether this information can help them identify better viewpoints. Mooney et al.\ investigated the perception of stress~\cite{mooney2024stress}, asking participants to determine which of two drawings had higher stress.
They found that the task was feasible above a certain threshold difference.

In 2D settings, Huang et al.\ demonstrated that optimising a combination of criteria can improve task performance~\cite{huang2013improving}.
Several methods have been developed to optimise multiple quality measures simultaneously, or to support this through interactive user involvement~\cite{sponemann2014evolutionary,Ryall97,Coleman96}.
Since layouts can either facilitate or hinder task performance, they also influence cognitive load~\cite{Yogh20,Huang09,Ware02}, which is another consideration for future work in our context.
Closely related to our approach of exploring quality measures through user preferences are those that invert the classical `algorithm first, evaluation second' paradigm, deriving insight from human-generated layouts instead~\cite{Kieffer16,purchase2012graph}.

In 3D graph visualisation, Joos et al.\ computed aesthetic measures for a large number of sampled viewpoints and visualised these in a VR environment, helping users identify potentially beneficial perspectives, which received positive feedback from domain experts~\cite{joos2023aesthetic}.
Similarly, Wageningen et al.\ investigated how 2D quality measures vary across viewpoints sampled on a sphere surrounding 3D graph layouts, and applied gradient descent methods to find views with the highest aesthetic values~\cite{an2024wagendingen,viewpoint2025wagendingen}.
The investigation of the impact of such measures in stereoscopic 3D is more complex than in 2D, as factors such as depth, distance, and perspective might influence the perception significantly, as e.g., for edge crossings~\cite{zhang2025investigatingcrossingperception3d}.
Drogemuller et al.~\cite{drogemuller2023aesthetics} studied edge curvature aesthetics in Lombardi-inspired 3D layouts within stereoscopic 3D, finding that straight edges outperformed curved ones.
Bennett et al.~and Ware et al.\ have also examined the perceptual foundations of graph visualisation, including links to Gestalt principles~\cite{Bennett07,Ware02}.

Although considerable research has addressed drawing principles for 2D graph visualisations, and some initial work has explored advantageous viewpoints in 3D drawings, there remains no empirical evaluation of user-preferred perspectives in 3D and whether these preferences can be characterised by existing aesthetic measures or require new ones.

\section{Methodology}
\label{sec:methodology}

To examine how viewers choose preferred viewpoints in 3D graph drawings, and to what extent these can be explained by aesthetic measures (see \autoref{sec:introduction}), we conducted an empirical user study and analysed the data through a structured pipeline (see \autoref{sec:results}).
Participants were asked to rotate a series of 3D graphs, and for each, to identify best and worst viewpoints.
In the following, we describe the dataset, the VR prototype, the study procedure, the aesthetic measures, and participant details.

\subsection{Stimuli}
\label{sec:methodology-data}

Drawing on experience from earlier VR studies, we aimed to keep headset time below 45 minutes.
With an average of one minute per graph plus setup and calibration, we decided on a sample of 36 graphs.
To support general and class-specific conclusions, we included graphs from four size classes (\GSize{S} (20 nodes), \GSize{M} (50), \GSize{L} (100), and \GSize{XL} (200)) and three layout types (energy-based (\GLayout{E}), layered (\GLayout{L}), and semantic (\GLayout{S}) types), aligned with those typically used in immersive settings~\cite{joos2025visual}.
With three graphs per size-layout combination, this yields 36 graph drawings.
Edge density varied between 0.75\% and 33.3\%, computed as $\frac{2m}{n(n-1)}$ for a graph $G = (V, E),\; |V| = n,\; |E| = m$.
\autoref{fig:perspective-sample-grid} shows a sample of these graphs, while a full documentation can be found in \autoref{sec:appendix-chosen-perspectives-spheres-all}.
Following, we describe each class and how the 3D graph drawings were created.

\paragraph*{Energy-Based Layouts (L-E)}
Energy-based algorithms are widely used for immersive graph visualisations, particularly when node positions lack semantic meaning, as many drawing algorithms natively support 3D.
For this layout type, we selected 12 graphs (three per size) and computed 3D layouts using the \textit{Stress Minimisation} algorithm from OGDF~\cite{chimani2013open}.
Graphs were drawn from the \textit{Rome}~\cite{dibattista1997an} dataset and \textit{SuiteSparse Matrix Collection}~\cite{davis2011the}, both standard benchmarks. 
To include sparser, structured samples, we created graphs with the OGDF Random Tree Generator.

\paragraph*{Layered Layouts (L-L)}
Layered graphs are relevant in cases with hierarchical structures or distinct node types~\cite{mcgee2019state}.
We followed a method from a recent VR study on layered graphs for generation~\cite{feyer20242D}.
We used two layers for \GSize{S} and \GSize{M} and three for \GSize{L} and \GSize{XL}.
Given a target size $n$, nodes were partitioned into $l$ layers with a maximum deviation of $\pm 10\%$ from $\frac{n}{l}$, ensuring a roughly even split with some perturbation.
Within each layer, edges were added using OGDF's \textit{Simple Connected Graph Generator}, followed by the \textit{FMMM} layout algorithm for 2D node placement.
Between adjacent layers, 5 to $0.7n$ inter-layer edges were added at random to produce varying densities.

\paragraph*{Semantic Layouts (L-S)}
A key advantage of 3D visualisations is the ability to place nodes at meaningful spatial, semantic locations.
A prominent case is brain activity networks, where nodes represent brain regions (centroids) and edges indicate functional correlations~\cite{jaeger2019challenges, pirch2021vrnetzer}.
These networks are undirected, fully connected, and weighted.
Applying a threshold removes weaker edges, and discarding the weights yields an unweighted graph that highlights strong correlations.
Region granularity is adjustable via brain atlases, offering flexibility in node and edge counts.
Given their frequent use in immersive contexts, meaningful spatial structure, and adaptability, we included such networks in the evaluation.
For \GSize{L} and \GSize{XL}, we used human fMRI data from the Human Connectome Project~\cite{tijhuis2025hcp, tijhuis2022zenodo} with the Schaefer (100 regions) and Brainnetome (246 downsampled to 200) atlases~\cite{schaefer2017local, fan2016the}.
For \GSize{S} and \GSize{M}, we used rat fMRI data~\cite{achard2023inter, becq2022zenodo} (53 regions, downsampled to 20 or 50).
Based on prior experience, edge thresholds were tuned to yield realistic connectivity for fMRI analysis (see \autoref{tbl:aesthetics-measure-table-all}).

\subsection{Application}
\label{sec:methodology-prototype}

\begin{figure}
    \centering
    \includegraphics[width=1\linewidth]{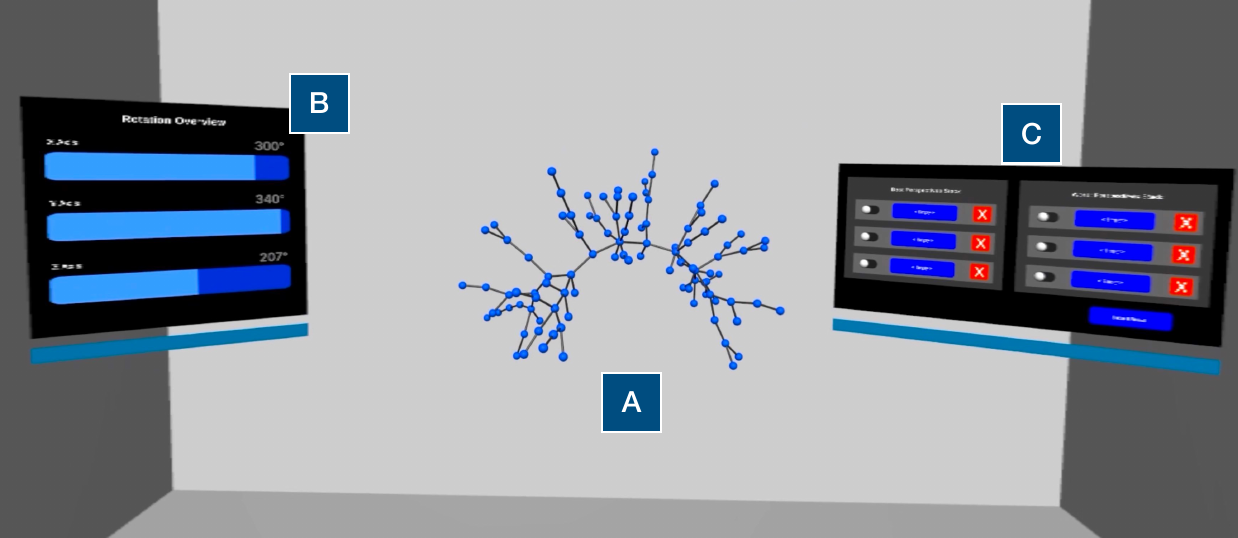}
    \caption{The VR study application, perceived and controlled using the Apple Vision Pro. The 3D graph drawing (A) can be rotated by a gaze and pinch interaction, or sliders (B). Users can store up to three preferred or disregarded perspectives (C).}
    \label{fig:study-application}
\end{figure}

We developed a VR application for the Apple Vision Pro (2024 release) using Unity.
The headset offered high resolution, substantial processing capabilities, and native hand- and gaze-based interaction, making it ideal for this evaluation.
The virtual environment consisted of a minimal room designed to provide spatial orientation while minimising distractions.
Participants experienced the environment while seated at a fixed position (only head movement was allowed).
Graphs with pre-computed 3D layouts were presented in stereoscopic space, using blue spheres for nodes and black tubes for edges -- a typical representation for graphs in immersive settings~\cite{joos2025visual} (see \autoref{fig:study-application}).
Users rotated the graph by directing their gaze at it and performing a pinch gesture (thumb and index finger together), applying the hand rotation to the graph.
Three sliders provided fine-grained rotation control around the X, Y, and Z axes.
An interactive panel allowed participants to store up to three viewpoints per category (best or worst), review them, and select one final view for each category using toggles.
Submissions were only enabled after both selections were made.
Upon submission, the selected views and associated metadata (e.g., task duration and graph position) were logged in a database.
Both user interface panels were movable within the 3D environment to avoid occlusion and support ergonomic interaction.
While graphs remained fixed during each task, participants could adjust the default placement of the visualisation between tasks by moving a bounding sphere.
Graph size was fixed, and full visibility was ensured at all times.

\subsection{Study Procedure}
\label{sec:methodology-study-procedure}

The study took place in a controlled university lab and lasted about one hour.
Participants received 15\,€ compensation.
After providing informed consent, participants watched a four-minute video explaining the setup, application, and task.
They were instructed to rotate each graph and identify both the \textit{best} and \textit{worst} viewing perspectives from their point of view.
Participants were informed that perspectives were not only to be evaluated based on visual appeal, but also on their ability to convey information relevant to task-solving.
After addressing any questions, the headset was fitted and calibrated.
A screen-mirroring setup allowed the instructor to monitor the session and offer guidance if needed.
Participants completed two trial tasks to familiarise themselves with the interface and determine a default graph position.
In the main study, they evaluated 36 graphs shown in random order.
Submissions were only possible after rotating the graph at least 180° along each axis ensuring a thorough exploration.
A soft time limit of one minute per task was set, which was not enforced, but participants were reminded if they exceeded it significantly.
Breaks were allowed at any time, and questions were addressed as they arose.
Audio was recorded, and participants were encouraged to verbalise their strategies.
At the end, participants completed a brief questionnaire on demographics, strategies, and their experience before leaving.

\subsection{Aesthetic Measures}
\label{sec:methodology-aesthetics}

Our analysis uses 21 aesthetic measures, mostly established 2D graph drawing metrics grounded in longstanding principles and empirical studies~\cite{mooney2024landscape,Bennett07,purchase2002metrics}.
Moreover, we include 3D-specific overlap measures and a novel metric for structural perceptibility in 3D.

\paragraph*{2D Measures}

To obtain 2D positions for aesthetic evaluation, we applied perspective projection to the 3D graph from a given viewpoint.
We used the Python library \texttt{GdMetriX}~\cite{nollenburg2024GdMetriX}, which implements all relevant 2D measures found in the literature and required for our analysis.
The following measures were calculated: \textit{edge crossings} \MeasureCR{}, \textit{stress} \MeasureST{}~\cite{welch2017measuring} (discrepancy of geometric and topological distances), \textit{crossing angular resolution} \MeasureCAR{} (how much deviate angles of crossing edges from an optimum), \textit{bounding box area} \MeasureAR{}, \textit{aspect ratio} \MeasureASP{}, \textit{node concentration} \MeasureCON{}~\cite{taylor2005applying} (estimating how (un)evenly distributed nodes are), \textit{node orthogonality} \MeasureNO{} (how `packed' are nodes on the grid), \textit{Gabriel ratio} \MeasureGR{}~\cite{mooney2024landscape} (measuring whether nodes are too close to edges), \textit{angular resolution} \MeasureANGR{}~\cite{purchase2002metrics} (the angle deviation of incident edges from an optimum), \textit{edge orthogonality} \MeasureEO{}~\cite{purchase2002metrics} (measuring how horizontal/vertical edges are), and \textit{edge length deviation} \MeasureELD{}~\cite{mooney2024landscape} (the mean deviation of edge lengths, aiming for uniformity).
To capture symmetry~\cite{meidiana2020quality}, we included three edge-based measures~\cite{klapaukh2018symmetry}: \textit{reflective} \MeasureESR{}, \textit{rotational} \MeasureESO{}, and \textit{translational} \MeasureEST{} symmetry, measuring whether a symmetry axis exists and quantifying rotational or translational invariance.

\paragraph*{3D Overlap}

Unlike in 2D, where edge crossings are the main form of occlusion, 3D graph drawings are more susceptible to various overlaps.
Prior work highlights their relevance for users~\cite{joos2023aesthetic}, yet most studies only consider node-node or edge-node overlaps, without accounting for the order or visual depth.
User feedback (see \autoref{sec:results-qualitative}) suggests that the direction of occlusion matters.
When a node appears in front of an edge, it can falsely imply a connection where none exists, making this type of overlap particularly misleading.
Conversely, when an edge passes in front of a node, the node often remains partially visible because edges are typically thinner.
However, this can still obscure important parts of the node and lead to misinterpretation, especially in dense regions.
Users also noted that the degree of overlap influences perception, not only their pure existence.
To reflect this, we include six overlap measures: \textit{node-node} \MeasureNNO{}, \textit{edge-node} \MeasureENO{}, and \textit{node-edge} \MeasureNEO{} count occurrences, while \textit{node-node overlap area} \MeasureNNOA{}, \textit{edge-node overlap area} \MeasureENOA{}, and \textit{node-edge overlap area} \MeasureNEOA{} quantify extent.
Node-node areas (circle-circle intersections) can be computed in closed form~\cite{WeissteinCircleCircleIntersection}.
Overlaps involving rectangles (edges) and circles (nodes) are more complicated and approximated via fine-grained rasterisation.

\paragraph*{Isometric Viewpoint Deviation}

While previous measures capture various aspects of viewpoint quality, they do not assess whether a graph's global structure is perceivable from a given perspective.
This aspect, although difficult to quantify, is crucial for understanding user preferences (see \autoref{sec:results-qualitative}).
It is not always clear what the overall structure of a graph is, or how to determine whether a viewpoint supports its perception, especially for arbitrary graphs with no prior semantic knowledge.
However, user feedback in our study provided valuable insight into how participants perceive structure.
Many described preferring a `diagonal view' that offers a balanced look along the graph's main axes, avoiding alignment between those axes and the view direction.
For most graphs, the perceived axes correspond to directions with high node spread, which can be estimated using principal component analysis (PCA).
Inspired by Farish's concept of \textit{isometric projections}~\cite{farish1822isometrical}, established in the 19th century, which are views in which the axes appear equally foreshortened (120° angle between each pair), we define a measure to quantify deviation from this ideal, taking into account whether PCA-derived axes meaningfully describe the graph's structure (\textit{anisotropy}).
Let the normalised view vector be $\boldsymbol{v} \in \mathbb{R}^3$, and the PCA of the node positions yield orthonormal eigenvectors $\boldsymbol{E} = [\boldsymbol{e}_1, \boldsymbol{e}_2, \boldsymbol{e}_3] \in \mathbb{R}^{3 \times 3}$ and corresponding eigenvalues $\boldsymbol{\lambda} = [\lambda_1, \lambda_2, \lambda_3]^\top$.
We define the \textit{Isometric Viewpoint Deviation} score as 
$
\mathrm{ISO} = 1 - \alpha \cdot \frac{\sigma_w}{\sigma_{\max}}
$ with 
\begin{itemize}
  \setlength\itemsep{1mm}
  \item $\boldsymbol{a} = \left|\boldsymbol{E}^\top \boldsymbol{v}\right|$: absolute projections on principal axes,
  \item $\boldsymbol{a}_{\text{norm}} = \boldsymbol{a} / \sum a_i$: normalised projection weights,
  \item $\mu = \sum w_i\, a_{\text{norm},i}$: weighted mean,
  \item $\sigma_w = \sqrt{\sum w_i\, \left(a_{\text{norm},i} - \mu\right)^2}$: weighted standard deviation of projections,
  \item $\alpha = \sigma(\boldsymbol{w}) / \sigma_{\max}$: normalised anisotropy factor,
  \item $\sigma_{\max} = 1/\sqrt{3}$: normalisation constant.
\end{itemize}

\noindent
The anisotropy factor $\alpha$ ensures the score only penalises unbalanced projections when the graph has meaningful, deviating main axes (i.e., uneven eigenvalues).
It is zero for isotropic graphs and one for maximally anisotropic ones.
This is combined with $\sigma_w$, which captures how unevenly the principal axes are represented in the current view, normalised by the worst-case imbalance.
The resulting score \MeasureISO{} ranges from 0 to 1, where 0 indicates highly imbalanced views of strongly structured graphs, and 1 corresponds to either balanced views or graphs lacking dominant directions.
While we acknowledge that there may be other, potentially better or more sophisticated measures to quantify the perception of main axes or to define \textit{structure}, both the tests and user-reported strategies suggest that this approach is worth incorporating into the analysis and can serve as a baseline for future methods.

\subsection{Participants}
\label{sec:methodology-participants}

We recruited 23 participants (9 female, 14 male) from among students and associates at our university to take part in the study.
The average age of participants was 25.96 years (SD = 3.94), with a median age of 25.
All participants reported normal or corrected-to-normal vision.
Corrected vision was achieved through contact lenses, as the VR device can not be used with glasses.
Participants rated their experience with graphs on a five-point Likert scale ranging from 1 (No Experience) to 5 (Expert Level Experience), resulting in a mean rating of 2.52.
Their experience with immersive technologies (such as augmented or virtual reality) was assessed using the same scale, with a mean rating of 2.27.
After completing the study, participants rated the task difficulty on a five-point Likert scale from 1 (Very Easy) to 5 (Very Difficult), yielding a mean difficulty rating of 2.17.

\section{Results}
\label{sec:results}

In the following, we present the qualitative feedback provided by participants, examine how their reported strategies align with their selected viewpoints, and analyse how these preferences vary.
We also evaluate the aesthetic measures applied in the study and assess the extent to which individual or combined measures explain users' choices.
Numbers in parentheses indicate how participants mentioned a strategy or observation.

\subsection{Qualitative Results}
\label{sec:results-qualitative}

During and after the study, participants described their strategies for identifying the most and least preferred viewpoints.
Across all graph types, a central theme was ensuring the overall structure was visible (\(n=17\)), even at the cost of some overlap or edge crossings.
`Diagonal' views on the graph were considered beneficial for this purpose (\(n=3\)).
Minimising node and edge overlap was frequently mentioned (\(n=5\)), with some participants focusing specifically on node-edge overlap (\(n=5\)), and one on edge-node overlap.
Others stressed the importance of distinguishing connected from unconnected nodes (\(n=1\)), and warned against heavy node overlap (\(n=3\)), especially in depth (\(n=1\)), where structure became hard to discern otherwise (\(n=2\)).
An observed strategy involved first finding a view revealing the global structure, then refining the view to reduce overlaps and crossings (\(n=3\)).
Generally, global understanding was prioritised over resolving all local ambiguities.
Participants identified several detrimental factors: edge crossings, compressed visualisation (especially into depth), misleading projections, and excessive flattening into 2D.
Wide layouts were preferred over tall ones, as they better matched the VR field of view.
Some noted that misleading visual cues, such as nodes occluding edges, suggesting false connections (\(n=4\)), were problematic, as were edges overlapping nodes (\(n=2\)).
However, one participant mentioned that partially occluded long edges were mostly acceptable, as their course could still be inferred (Gestalt principle of continuity).
Partial visibility of nodes or edges was sometimes sufficient for comprehension, while a larger projected area appeared to support understanding (\(n=2\)).

For layered graphs, a `tilted side view' was frequently communicated as most effective (\(n=9\)), while `orthogonal' views often led to overlap and were less helpful (\(n=7\)).
Similar comments applied to semantically laid-out graphs, especially where parallel edges aligned with the view axis (\(n=3\)).
Energy-based layouts prompted more diverse strategies.
For tree-like structures, views that aligned the main branch with the view axis were discouraged (\(n=2\)), with diagonal views again preferred (\(n=1\)).
In multi-component graphs (e.g., E-9, see \autoref{fig:appendix-b-e-9}), overlap among components was a concern (\(n=4\)), and placing the more complex part in front improved clarity (\(n=3\)).
The trade-off between revealing structure and reducing overlap was especially apparent for graph E-2.
Participants noted that a slightly diagonal, non-planar view allowed the structure to be perceived, even at the cost of increased overlap (\(n=3\)).
For E-6, the optimal view appeared task-dependent: diagonal views helped expose internal structure (\(n=3\)), while frontal or side views reduced overlap but obscured the overall form (\(n=2\)).
Individual users sought a view that would reveal the full structure in a way that could be reconstructed by others (\(n=1\)), attempted to identify principal components (\(n=1\)), or aimed to balance visibility with occlusion (\(n=1\)).
While selecting the best view was considered challenging in some cases, participants found the 3D environment intuitive and helpful for structural insight.
Three participants compared the experience to viewing molecular models, and two noted that 3D perspectives revealed structures that would be difficult to interpret in 2D without stereoscopic cues.

\subsection{User-Chosen Perspectives}
\label{sec:results-chosen-perspectives}

\begin{figure}[p]
    \centering
    \includegraphics[width=1\linewidth]{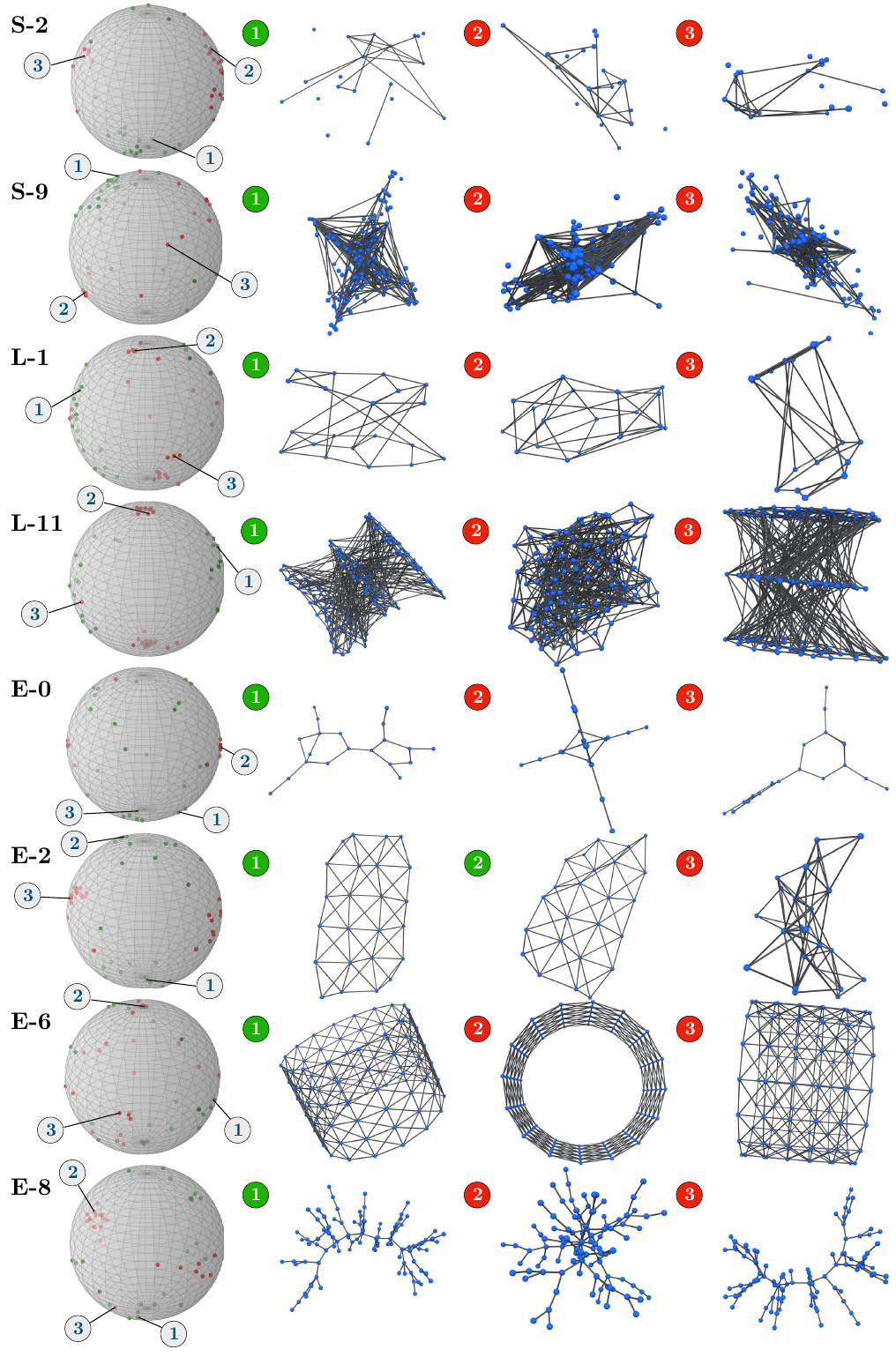}
    \caption{Example of eight graphs (of 36, reflecting the different types) showing the distribution of selected perspectives (sphere surface) along with three perspective projections (either \textit{best} (green) or \textit{worst} (red)), discussed in \autoref{sec:results-chosen-perspectives}. The complete set of perspectives is shown in \autoref{sec:appendix-chosen-perspectives-spheres-all}.}
    \label{fig:perspective-sample-grid}
\end{figure}

In the following, we discuss representative user-selected perspectives and their distributions for a sample of eight graphs (see \autoref{fig:perspective-sample-grid}).
All selected perspectives across users and graph visualisations are shown in \autoref{sec:appendix-chosen-perspectives-spheres-all}.

For \GLayout{S} graphs, users made relatively consistent selections.
In smaller, less dense graphs (e.g., \texttt{S-2}), the `optimal' perspectives spread out the graph, minimising occlusion.
In `worst' views, the graph appears compact with heavy overlap.
In more complex graphs (e.g., \texttt{S-9}), `positive' perspectives reveal structure along distinct axes.
In contrast, `poor' views align these axes with the viewing direction, causing occlusion, which is consistent with reported strategies.
\GLayout{L} graphs (e.g., \texttt{L-1}, \texttt{L-11}) show similar patterns.
`Best' perspectives expose the layered structure, typically via a slightly angled view, matching participant strategies.
In `worst' views, a front-on view of one layer hides others, leading to strong overlap.
Some users instead viewed layers orthogonally from the side, causing intra-layer node overlap.
This was more common in simpler graphs (e.g., \texttt{L-1}) and rare for more complex ones (e.g., \texttt{L-11}).
In \GLayout{E} graphs, perspective selections varied more due to differing topologies despite identical layout algorithms.
Structure remained a key factor.
In \texttt{E-0}, diverse `good' perspectives still conveyed the structure well.
For `negative' views, participants often selected misleading angles, e.g., cross-shaped (\texttt{2}) or partially obscured (\texttt{3}) structures.
In \texttt{E-2}, a clear 2D projection (\texttt{1}) was rarely chosen.
Instead, users preferred a less optimal projection (\texttt{2}) in terms of crossings but clearer 3D form, as supported by verbal feedback.
`Worst' perspectives failed to show the structure and caused major overlap.
In \texttt{E-6}, resembling a cylinder, users varied more.
Many chose a top-down, angled view (\texttt{1}) as `optimal'.
Others preferred (\texttt{2}), which had fewer crossings but poorer structural clarity.
Those who did not select it as `worst' chose a side-on orthogonal view with severe occlusion.
In tree-like graphs (e.g., \texttt{E-8}), `good' perspectives revealed the trunk clearly (\texttt{1}), while `bad' views rotated it into the image plane (\texttt{2}).
A few alternatives showed overlapping side branches (\texttt{3}).

Overall, the strategies communicated by the participants are clearly reflected in the results.
Some graphs prompted highly consistent choices, while others showed broader variation.

\begin{table}[b!]
    \centering
    \caption{Mean aesthetic values (\textit{best}~$\uparrow$, \textit{worst}~$\downarrow$) for all graphs (\texttt{All}), layout classes \GLayout{S} (\textit{semantic}), \GLayout{L} (\textit{layered}), and \GLayout{E} (\textit{energy-based}), and size classes \GSize{S} (20 nodes), \GSize{M} (50), \GSize{L} (100), and \GSize{XL} (200). In addition to 21 aesthetic measures, combined scores (RdYlGr colour scale) using logistic regression (\MeasureCLR{}) and sequential quadratic programming (\MeasureCSQP{}) are shown (5 aesthetics each).}
    \includegraphics[width=\linewidth]{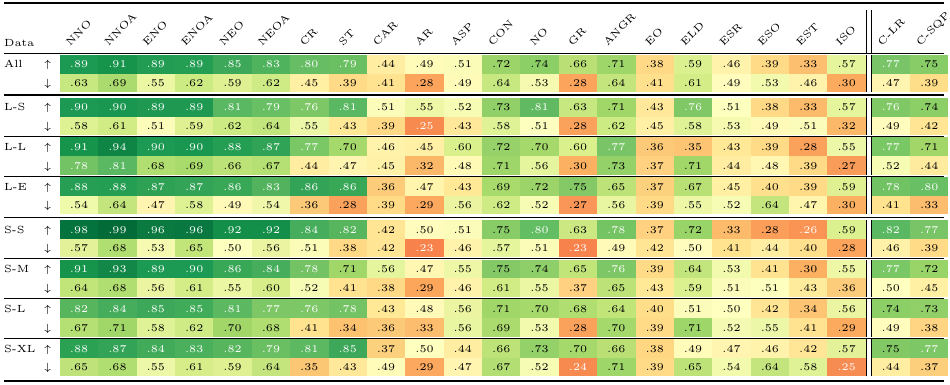}
    \vspace{-1mm}
    \label{tbl:aesthetic-measures-aggregated}
\end{table}

\subsection{Aesthetic Measures}
\label{sec:results-aesthetics}

\begin{table}[h!]
    \centering
        \caption{For all 36 graphs, we computed the mean values of 21 aesthetic measures and their linear combinations weighted by logistic regression (\texttt{C-LR}) and sequential quadratic programming (\texttt{C-SQP}).
Values are shown for participant-favoured ($\uparrow$) and -disliked ($\downarrow$) perspectives. The table also lists the number of nodes $|V|$ and edges $|E|$ per graph and is grouped by graph type: \textit{semantic} (\texttt{S}), \textit{layered} (\texttt{L}), and \textit{energy-based} (\texttt{E}).}
    \includegraphics[width=1\linewidth]{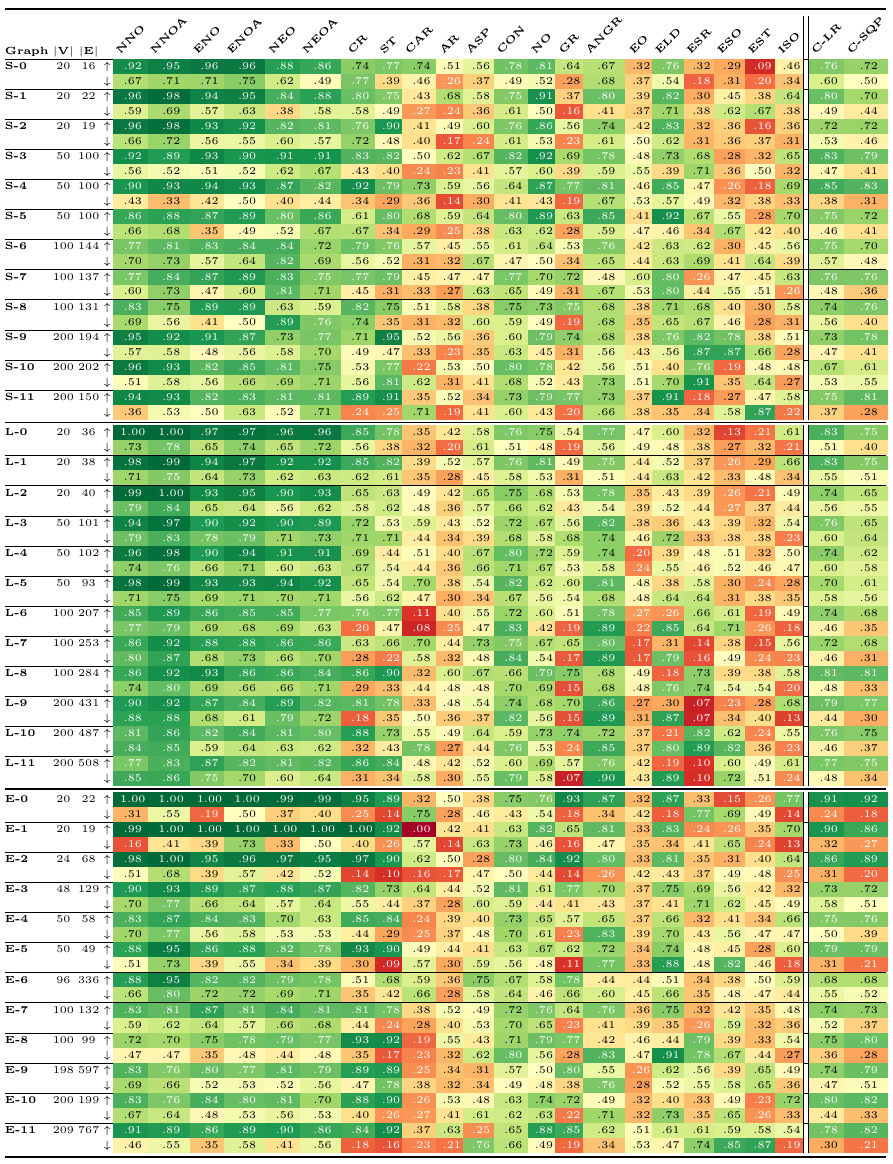}
    \label{tbl:aesthetics-measure-table-all}
    \vspace*{-2mm}
\end{table}

Before calculating the aesthetic measures, we converted each chosen perspective's graph position and rotation, along with the user position and rotation, into a comparable view vector representation.
We then computed 2D graph projections using a perspective projection matching the VR headset's camera parameters.
In addition to node and edge positions, the projected dimensions of the representation objects (circles and rectangles) were calculated.
To ensure comparability, all measures were adapted such that 0 indicates lowest and 1 highest quality.
For each selected perspective (best and worst), we computed the aesthetic measures introduced in \autoref{sec:methodology-aesthetics}.
Additionally, we uniformly sampled 5000 perspectives per graph using \textit{Fibonacci Lattice} and calculated all aesthetic measures for these as well.
This enabled us to determine the full range of achievable values and normalise the selected perspectives accordingly.
The resulting scores indicate where each selected perspective falls within this range.
\autoref{fig:histograms-aesthetic-measures} presents score distributions across all measures and selected perspectives, and \autoref{tbl:aesthetic-measures-aggregated} summarises mean scores across all data, layout types, and sizes (see \autoref{tbl:aesthetics-measure-table-all} for graph-level results).
The findings suggest that several measures, i.e., \MeasureCAR{}, \MeasureASP{}, \MeasureCON{}, \MeasureANGR{}, \MeasureEO{}, \MeasureELD{}, \MeasureESR{}, \MeasureESO{}, and \MeasureEST{}, do not characterise user choices.
In contrast, the six overlap measures, along with \MeasureCR{}, \MeasureST{}, \MeasureNO{}, \MeasureGR{}, and \MeasureISO{}, appear more relevant, as they differentiate best and worst perspectives to a certain extent.
A user-centric analysis of the aesthetic measure results did not reveal any outliers or substantial user-specific anomalies.
A correlation analysis (\autoref{fig:pca-corr-analysis}, right) shows that the overlap measures are highly correlated (particularly overlap counts with corresponding areas, which is not surprising).
Other notable correlations include those between \MeasureNO{} and \MeasureAR{}, and among \MeasureCR{}, \MeasureST{}, and \MeasureGR{}.
In \autoref{sec:results-combinations}, we further examine how well individual and combined measures explain user preferences.

\subsection{Combinations of Aesthetics}
\label{sec:results-combinations}
\begin{figure}[b!]
    \begin{subfigure}{0.51\textwidth}
        \centering
        \includegraphics[width=\linewidth]{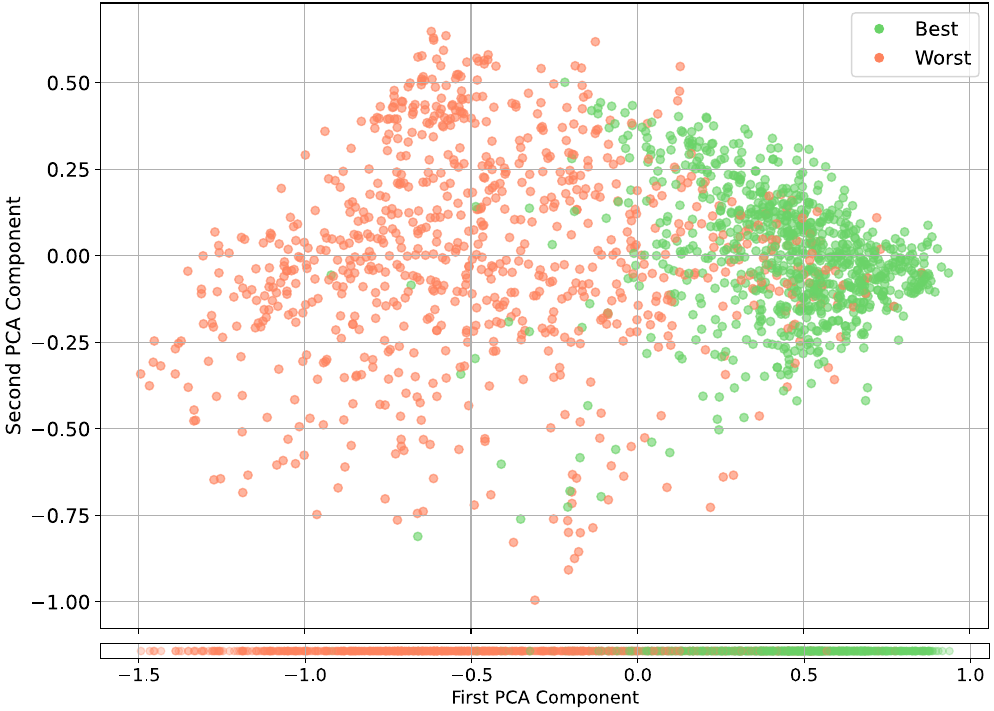}
    \end{subfigure}%
    \hfill%
    \begin{subfigure}{0.42\textwidth}
        \centering
        \includegraphics[width=\linewidth]{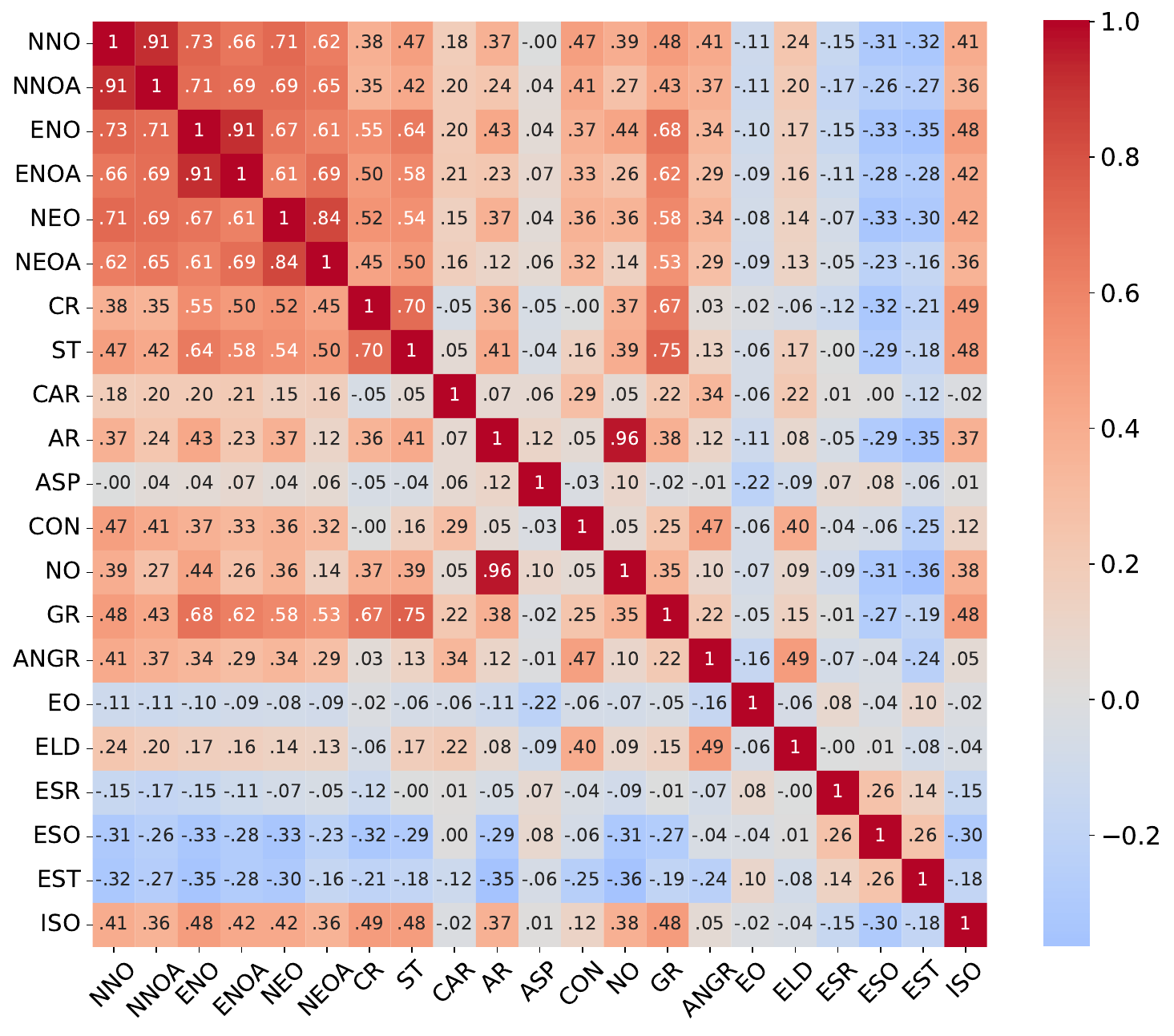}
    \end{subfigure}
    \caption{The first two components of a PCA of the \textit{best} (green) and \textit{worst} (red) user-chosen perspectives (left), in total $2 \times 36 \times 23 = 1.656$ samples with 21 measures each. The projection on the first component (left, below) already shows a good separation of the data points. A correlation analysis indicates that some aesthetic measures are highly correlated (especially the overlap measures \textsf{NN*}, \textsf{EN*}, and \textsf{NE*}), while others are not (right).}
    \label{fig:pca-corr-analysis}
\end{figure}
\begin{figure}
    \centering
    \includegraphics[width=\linewidth]{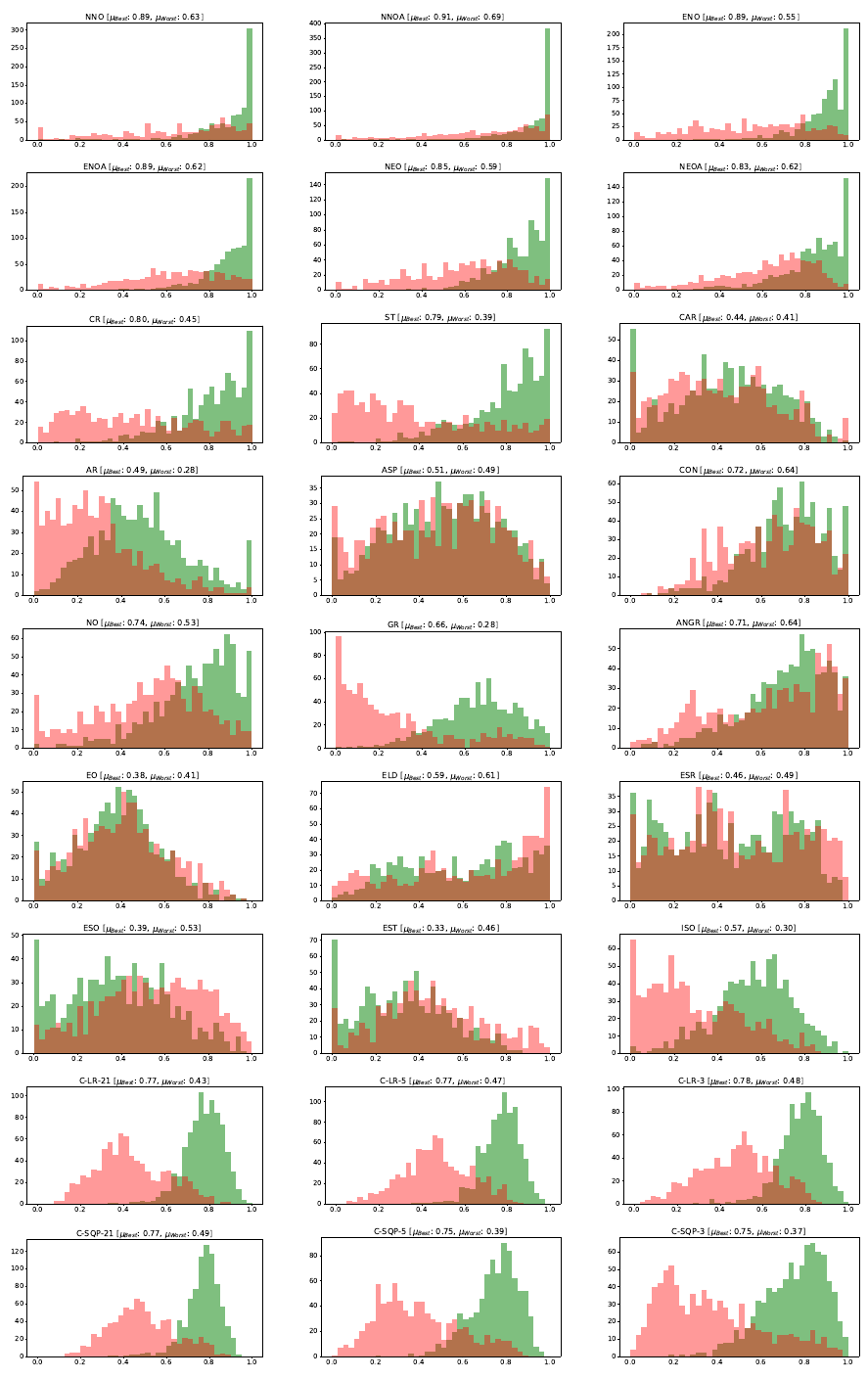}
    \caption{The distributions of values (\textit{best}: green, \textit{worst}: red) for 21 aesthetic measures (top seven rows) and combinations of these (\texttt{LR} and \texttt{SQP}) with 21, 5, or 3 measures each (last two rows).}
    \label{fig:histograms-aesthetic-measures}
\end{figure}
Based on the calculated aesthetic measures, we investigate how well they align with user judgements and whether combined measures can better reflect their importance.
A combined measure approximating user choices could be valuable in supporting the automated selection of \textit{optimal} viewpoints.
To get an initial impression of the data interdependences, we performed a PCA on the study results (see \autoref{fig:pca-corr-analysis}, left).
Interestingly, although not designed for this purpose, the first component already provides a good separation between best and worst perspectives.
Since each PCA component represents a weighted linear combination of the original features (here, aesthetic measures), we were motivated to explore interpretable, linear combinations of aesthetics that match the collected data.
Given these first insights, and considering both the interpretability and the typically lower risk of overfitting associated with linear models -- particularly in settings with limited data -- we aimed to find `optimal' linear aesthetic combinations corresponding to the collected data.
We pursued two different approaches for finding such combinations.
The first treats the task as a binary classification problem, using \textit{logistic regression} (LR) to distinguish best from worst views.
The resulting coefficients can be interpreted as importance weights.
However, this approach aims only to classify scores as above or below 0.5, not necessarily to maximise separation.
For clearer interpretability, assigning best views scores close to 1 and worst close to 0, we pursued an approach based on \textit{sequential quadratic programming} (SQP):
Given aesthetic measures $\{A_1,...,A_m\}$ with $A_i: p \rightarrow v$, where $p \in P$ is a perspective and $0 \leq v \leq 1$, and equally sized classes $P_{best}$ and $P_{worst}$, we optimise:
$
\left(\sum^n_p\sum^m_a w_a*A_a(P_{{best}_p})\right) - \left(\sum^n_p\sum^m_a w_a*A_a(P_{{worst}_p})\right)
$
This could be posed as a linear programme with a convex constraint ($w_a \geq 0$, $\sum w_a = 1$), but such formulations tend to create degenerate solutions with all weights assigned to a single measure.
Hence, we instead constrain the weight vector to satisfy $\|\vec{w}\|_2 = 1$ and $w_a \geq 0$, allowing more balanced distributions.
This constraint requires a non-linear solving approach, such as SQP, which handles the constraints efficiently and yields optimal weights maximising the separation between best and worst perspectives.

The resulting importance coefficients are shown in \autoref{tbl:combinations-importances}.
In addition to computing importances across the entire dataset, we also report results per layout and graph size, and under constraints limiting combinations to five or three measures, to identify the most influential ones.
Across conditions, the most relevant measures include the overlap measures, \MeasureCR{}, \MeasureST{}, \MeasureGR{}, and \MeasureISO{}.
Although we expected overlap area to be more important than count, this is not reflected in the results.
Interestingly, \MeasureENO{} is typically more relevant than \MeasureNEO{}, while \MeasureNNO{} appears less expressive.
For \GLayout{S} graphs, overlaps, particularly \MeasureNNO{}, along with \MeasureCR{}, \MeasureST{}, \MeasureGR{}, and \MeasureAR{} are most relevant, while \MeasureISO{} is less.
In \GLayout{L} graphs, \MeasureNEO{} dominates among overlap measures, with the usual others and especially \MeasureISO{} being important.
\GLayout{E} graphs show similar trends, though \MeasureENO{} is more prominent, while \MeasureISO{} is often excluded.
Regarding graph size, smaller graphs appear more affected by overlaps (including \MeasureNNO{}), whereas larger ones place more importance on \MeasureST{} and \MeasureGR{}.
Notably, the SQP method tends to favour measures like \MeasureST{} and \MeasureGR{}, where best and worst scores are spread out more, even if some worst views receive high scores.
In contrast, LR prefers measures that more clearly separate distributions, such as \MeasureISO{} and the overlap measures, resulting in better classifications (see \autoref{fig:histograms-aesthetic-measures}).
For comparison with individual aesthetics, we computed linear combinations using weights from both \MeasureCLR{} (logistic regression) and \MeasureCSQP{} (sequential quadratic programming).
The results appear to the right in all tables and in \autoref{fig:histograms-aesthetic-measures}.

\begin{table}[t!]
  \caption{The relative importance of the aesthetic measures when combined to express the perspective choices of users. We applied logistic regression (\texttt{LR}) and sequential quadratic programming (\texttt{SQP}) to find the most relevant aesthetic measures when using 21, 5, and 3 measures ($|A|$, i.e.\ shown as indicated measure cells per each row). We investigated the importance for all data at once (\texttt{All}), the three layout classes (\GLayout{S}, \GLayout{L}, \GLayout{E}), and the four graph sizes (\GSize{S}, \GSize{M}, \GSize{L}, \GSize{XL}).}
  \label{tbl:combinations-importances}
  \vspace{-2mm}
  \includegraphics[width=1.0\linewidth]{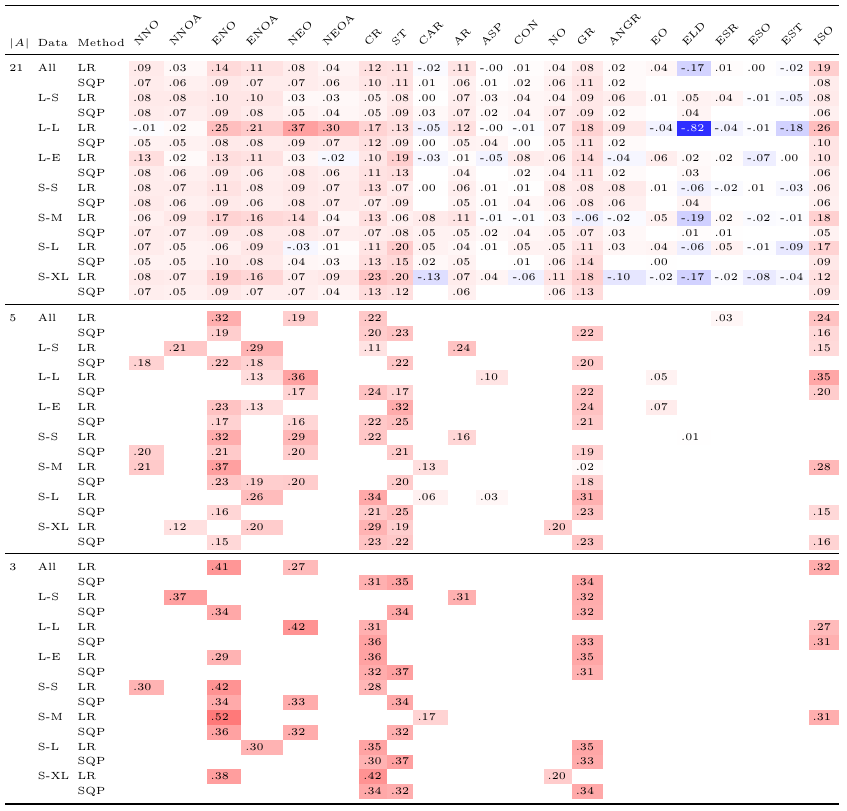}
  \vspace{-5mm}
\end{table}

\section{Discussion}
\label{sec:discussion}

Our analysis of user-selected viewpoints revealed several user strategies and demonstrated how their choices interrelate with aesthetic measures.
Notably, preferences varied across graph types and sizes, often reflecting trade-offs between structural clarity and local features.
This supports the view that users do not optimise for a single aesthetic, but rather negotiate among multiple perceptual cues, many of which are not captured by traditional 2D aesthetics.

The strongest 2D aesthetic predictors of viewpoint preference were \MeasureCR{}, \MeasureST{}, and \MeasureGR{}, aligning with prior findings in 2D~\cite{ChimaniPoster, mooney2024stress}, and extending their validity into S3D.
However, 3D-specific measures, such as \MeasureISO{} and the overlap aesthetics, proved highly relevant.
The prominence of \MeasureENO{} over \MeasureNNO{} is particularly noteworthy.
While one might expect node-node occlusions to be more perceptually disruptive, the results indicate that edge-node intersections may be more misleading.
Node-edge occlusions appeared less impactful, possibly due to the Gestalt principle of continuity.
Although overlap area was computed as a potentially more accurate representation, it showed no advantage over overlap counts.
This may reflect that users prefer views in which objects are clearly distinguishable, as indicated by the relevance of \MeasureGR{} penalising nodes and edges in close proximity.
Another key insight lies in the differing impact of layout type.
Layered and semantic layouts led to more consistent viewpoint choices than energy-based ones.
In semantic graphs, where node positions carry intrinsic meaning, participants prioritised preserving perceived structure.
This was reflected in the strong relevance of \MeasureISO{} for semantic and layered layouts, where users appeared to prefer perspectives maintaining orthogonal balance or offering diagonal overviews.
In contrast, energy-based graphs, which showed greater topological variation in our dataset, elicited more diverse and individual strategies, with user preferences often diverging even within the same graph.
These cases suggest that, in the absence of salient or meaningful structure, users default to evaluating local features such as component separation, edge crossings, or occlusions.

One somewhat surprising result was the limited explanatory power of several aesthetics such as \MeasureANGR{}, \MeasureASP{}, and symmetry measures.
Although considered beneficial in 2D, these did not significantly distinguish best and worst views in our study.
One explanation may be that such aesthetics target fine-grained aspects of local visual quality, while participants were more influenced by coarse, global cues, such as whether the overall structure `popped out' in space, or whether the orientation supported mental model construction.
Another possibility is that these aesthetics are harder to judge consistently in stereoscopic views, where depth cues from parallax and motion influence perception in ways not captured by 2D projections.

PCA and correlation analysis revealed that many aesthetic measures are indeed redundant, especially among the overlap measures.
Thus, a small number of well-chosen features can explain much of the variance in user preferences.
Our comparison of linear combination methods (logistic regression and sequential quadratic programming) confirms that different optimisation objectives yield subtly different rankings, though both approaches consistently identify the most predictive measures.
Combinations incorporating \MeasureST{}, \MeasureGR{}, \MeasureCR{}, \MeasureENO{}, and \MeasureISO{} consistently achieved the best separation between preferred and unfavoured viewpoints.

Beyond the numerical results, qualitative feedback reinforced the value of 3D interaction and stereoscopic cues in disambiguating structure, particularly in dense or ambiguous graphs.
Interestingly, user choices were largely consistent, with no substantial differences between novices and more experienced users.
This reinforces the value of identifying general principles for viewpoint preference, as we aimed to do in this work.

Despite careful design, this study has limitations.
Although we included a broad sample of graphs and layouts, results may differ for other or more specific datasets.
Moreover, our participants were primarily novices and generalisation to expert users or larger populations may require further study.
While we focused on user \textit{preference}, the relationship to task-solving performance remains unclear.
Thus, future work could build on previously introduced methods to identify aesthetic-optimal viewpoints, incorporating our findings on preference-driven combinations to investigate their effects on task performance.
Moreover, our analysis has so far treated each viewpoint as static and independent, whereas in real-world use, users navigate through sequences of views.
Studying viewpoint transitions, stability, and the temporal dynamics of understanding could offer insights into how users mentally assemble spatial structures over time.
Furthermore, based on our tests with the \MeasureISO{} measure, we see potential for developing new, 3D-specific aesthetic measures that reflect how users perceive graphs in a S3D setup.
Using our openly accessible dataset, such measures can be evaluated against the user preferences collected in this study.

\section{Conclusion}
\label{sec:conclusion}

In this work, we conducted a comprehensive study with 23 participants on viewpoint preference in 3D graph visualisation. 
Participants selected their most and least preferred viewpoints across 36 graphs varying in layout, size, and topology, and shared their selection strategies. 
Our findings show that, in addition to classical 2D aesthetic measures such as \textit{Stress}, \textit{Crossings}, and the \textit{Gabriel Ratio}, 3D-specific measures, especially \textit{Edge-Node Overlap} and overall structure-capturing methods like our proposed \textit{Isometric Viewpoint Deviation}, are highly reflecting user preference.
Participants were often willing to tolerate more overlap and crossings when a viewpoint conveyed the global structure more clearly.
Our analysis of combined aesthetics using logistic regression and sequential quadratic programming underlines these results, showing that a small set of measures characterises user choices to a high degree. 
However, the results indicate the need to adapt and extend traditional quality measures for the 3D context.
Beyond our empirical findings, the open-access dataset provides a foundation for future research on viewpoint evaluation and optimisation in immersive graph analysis.

\bibliography{bibliography}

\clearpage

\appendix

\setcounter{figure}{0}
\renewcommand{\thefigure}{A\arabic{figure}}

\input{appendix-b}

\end{document}

%% file: appendix-b.tex
\section{Chosen Perspective Projections and Distributions}
\label{sec:appendix-chosen-perspectives-spheres-all}
\vspace{5mm}

\begin{figure}[h!]
    \centering
    \includegraphics[width=1\linewidth]{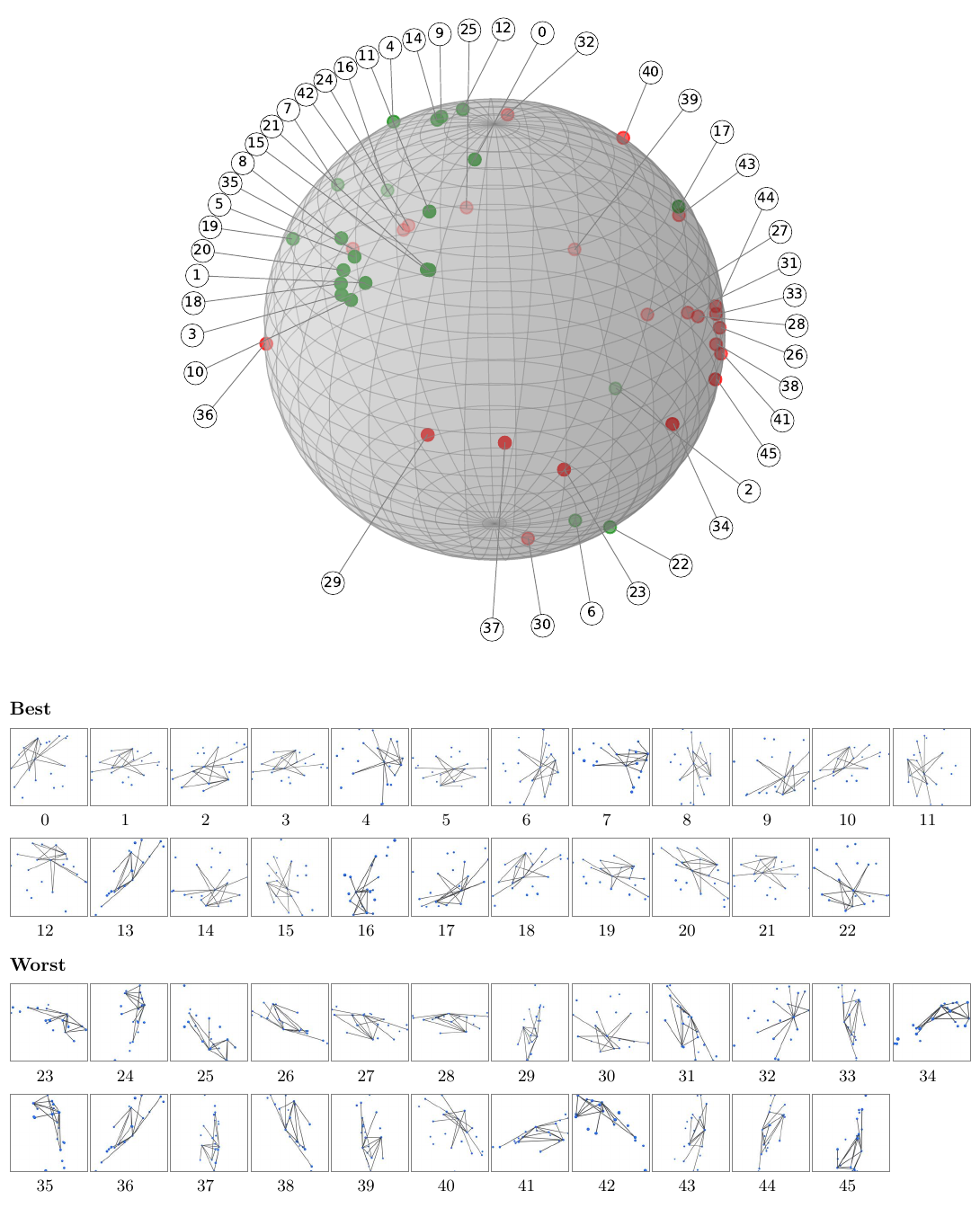}
    \caption{The distribution of all perspectives selected by users as \textit{best} (green) and \textit{worst} (red) mapped on a sphere surface for graph \textbf{S-0} (sem\_0\_20\_16) with $|V| = 20, \; |E| = 16$.}
\end{figure}

\begin{figure}
    \centering
    \includegraphics[width=1\linewidth]{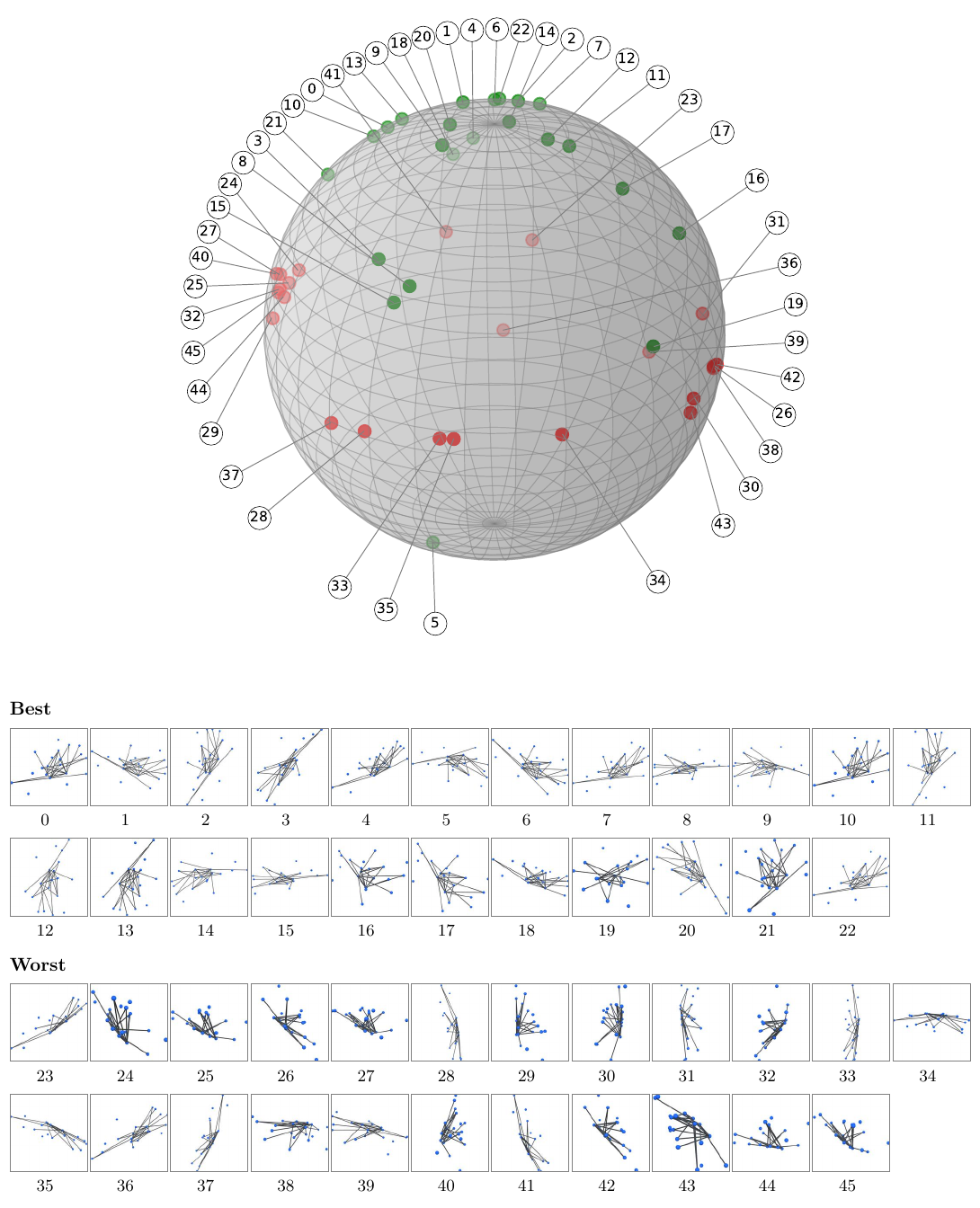}
    \caption{The distribution of all perspectives selected by users as \textit{best} (green) and \textit{worst} (red) mapped on a sphere surface for graph \textbf{S-1} (sem\_1\_20\_22) with $|V| = 20, \; |E| = 22$.}
\end{figure}

\begin{figure}
    \centering
    \includegraphics[width=1\linewidth]{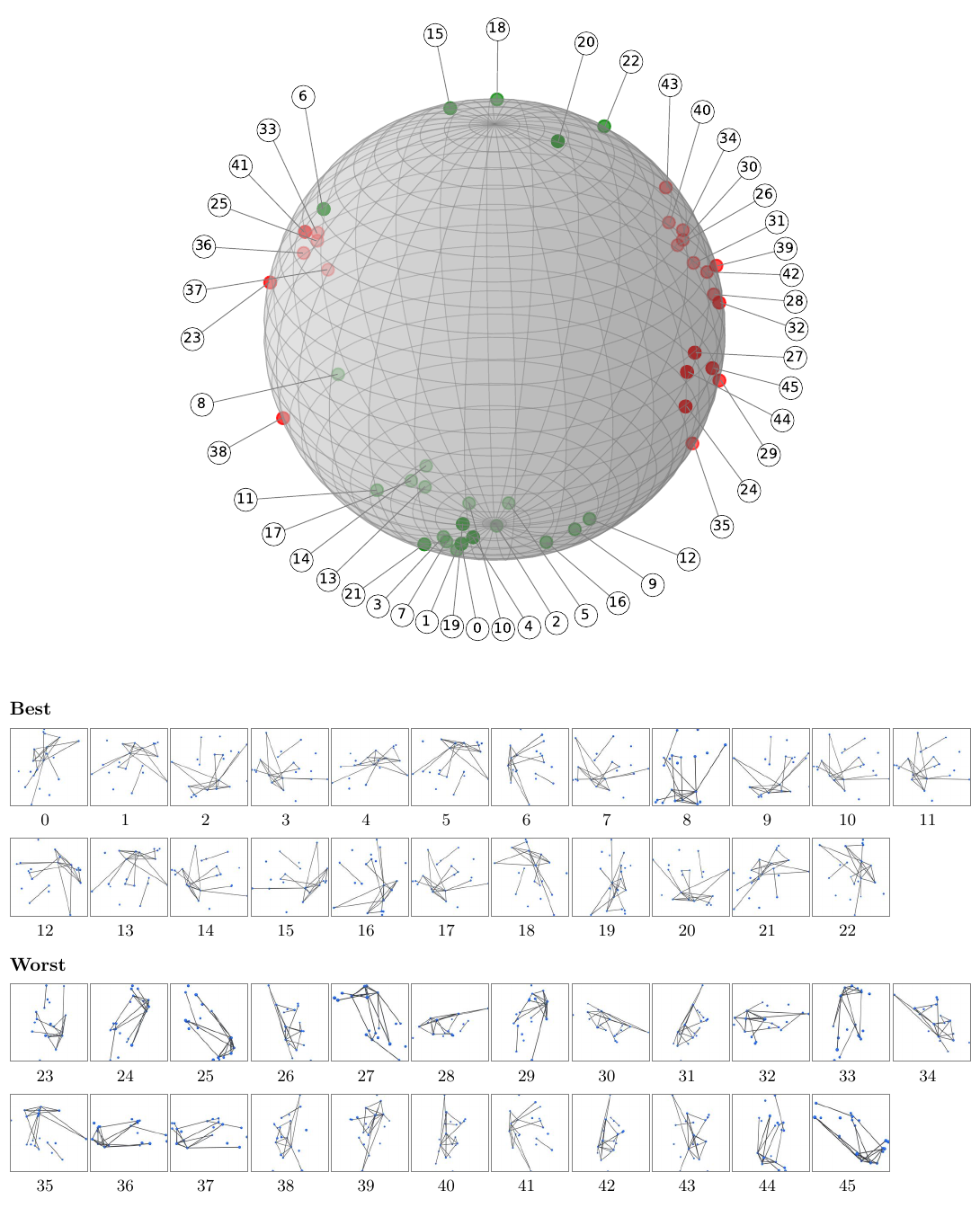}
    \caption{The distribution of all perspectives selected by users as \textit{best} (green) and \textit{worst} (red) mapped on a sphere surface for graph \textbf{S-2} (sem\_2\_20\_19) with $|V| = 20, \; |E| = 19$.}
\end{figure}

\begin{figure}
    \centering
    \includegraphics[width=1\linewidth]{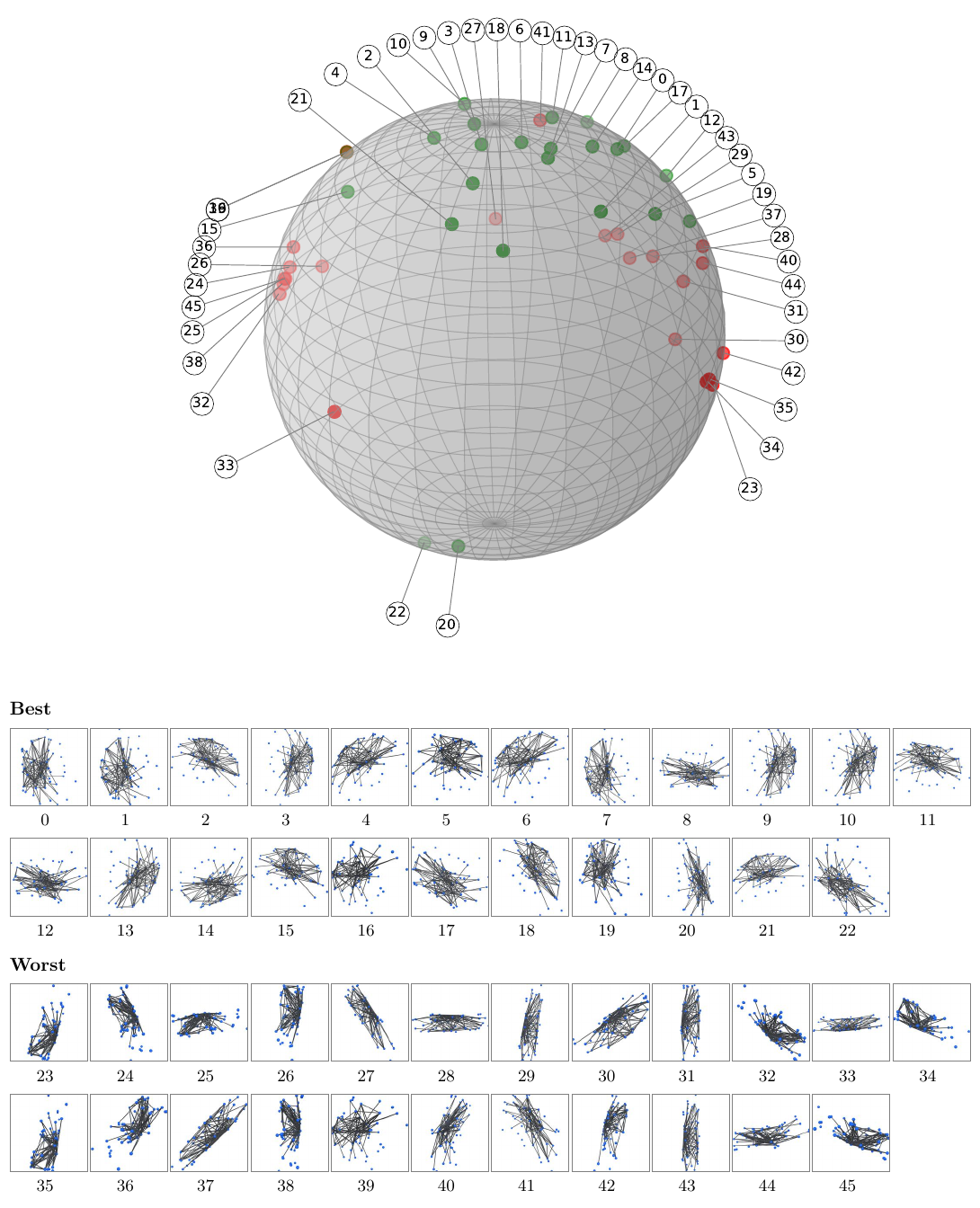}
    \caption{The distribution of all perspectives selected by users as \textit{best} (green) and \textit{worst} (red) mapped on a sphere surface for graph \textbf{S-3} (sem\_0\_50\_100) with $|V| = 50, \; |E| = 100$.}
\end{figure}

\begin{figure}
    \centering
    \includegraphics[width=1\linewidth]{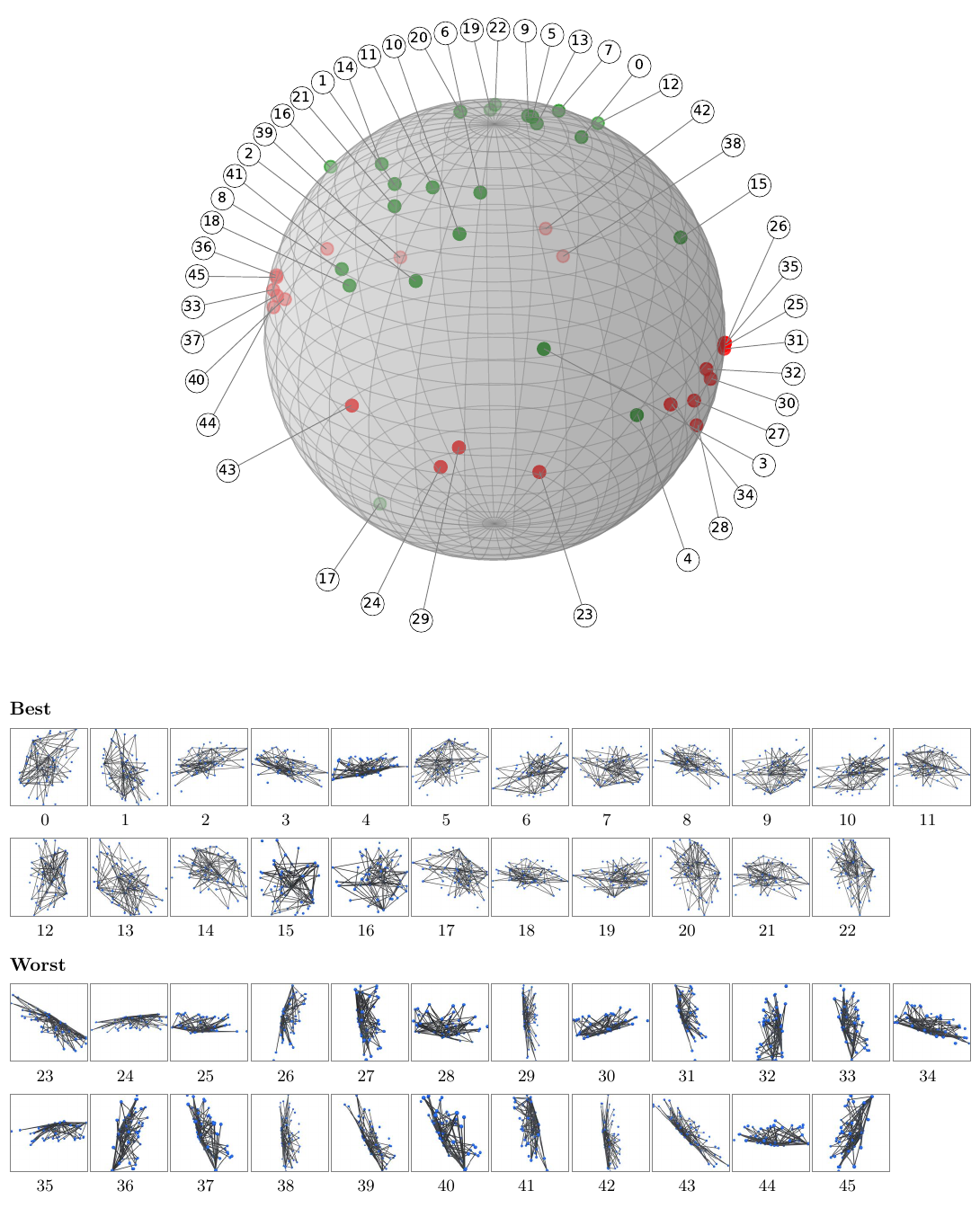}
    \caption{The distribution of all perspectives selected by users as \textit{best} (green) and \textit{worst} (red) mapped on a sphere surface for graph \textbf{S-4} (sem\_1\_50\_100) with $|V| = 50, \; |E| = 100$.}
\end{figure}

\begin{figure}
    \centering
    \includegraphics[width=1\linewidth]{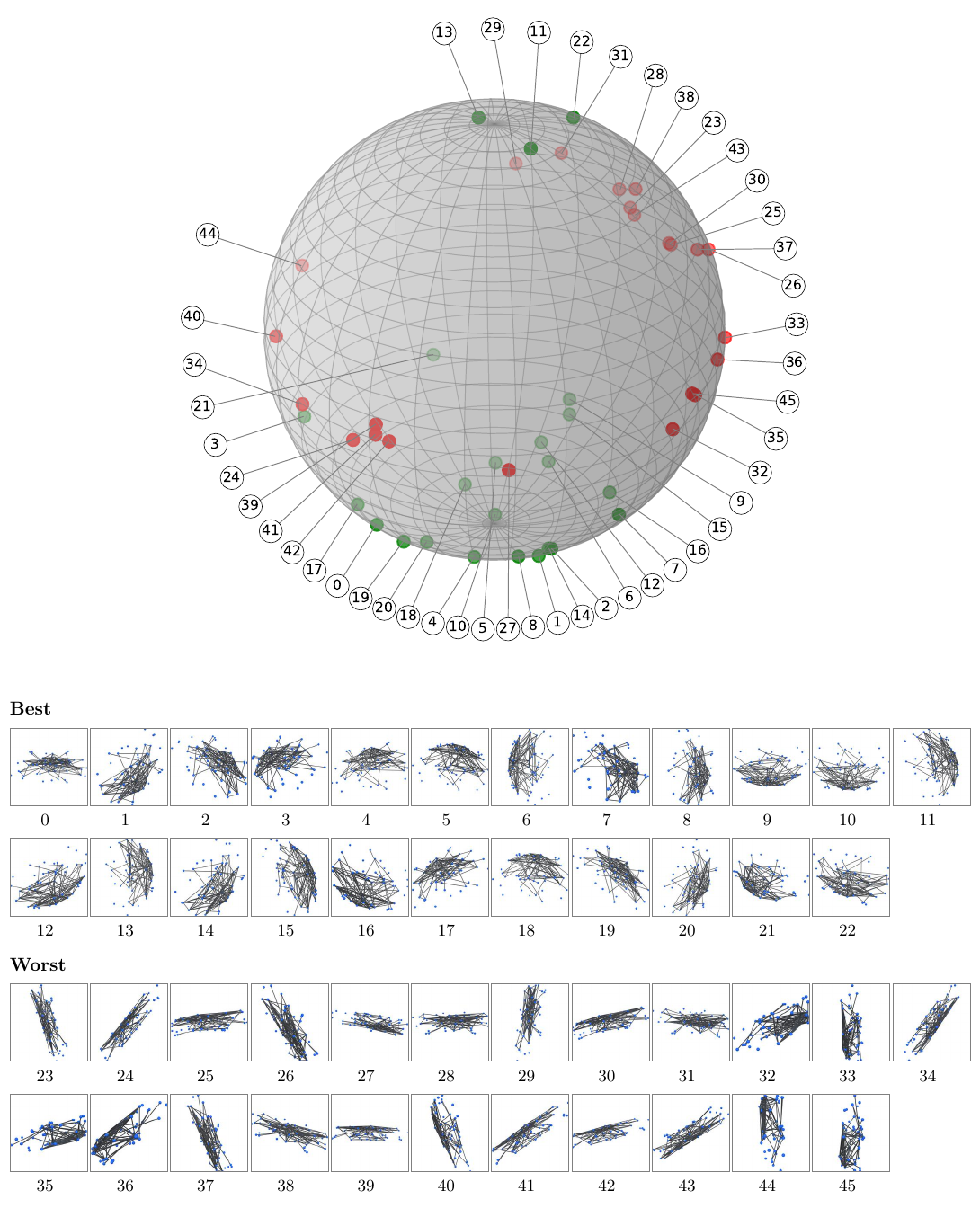}
    \caption{The distribution of all perspectives selected by users as \textit{best} (green) and \textit{worst} (red) mapped on a sphere surface for graph \textbf{S-5} (sem\_2\_50\_100) with $|V| = 50, \; |E| = 100$.}
\end{figure}

\begin{figure}
    \centering
    \includegraphics[width=1\linewidth]{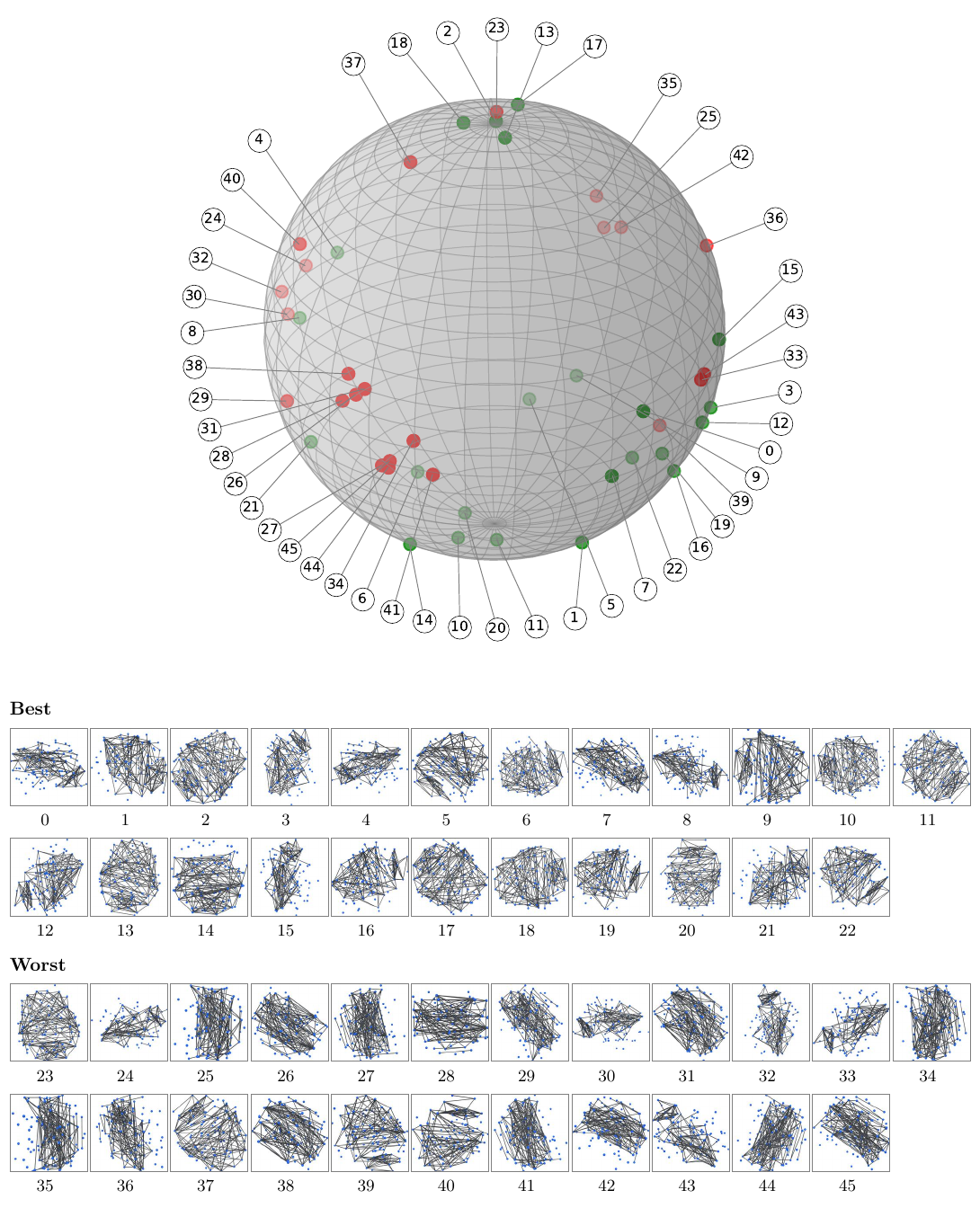}
    \caption{The distribution of all perspectives selected by users as \textit{best} (green) and \textit{worst} (red) mapped on a sphere surface for graph \textbf{S-6} (sem\_2\_100\_144) with $|V| = 100, \; |E| = 144$.}
\end{figure}

\begin{figure}
    \centering
    \includegraphics[width=1\linewidth]{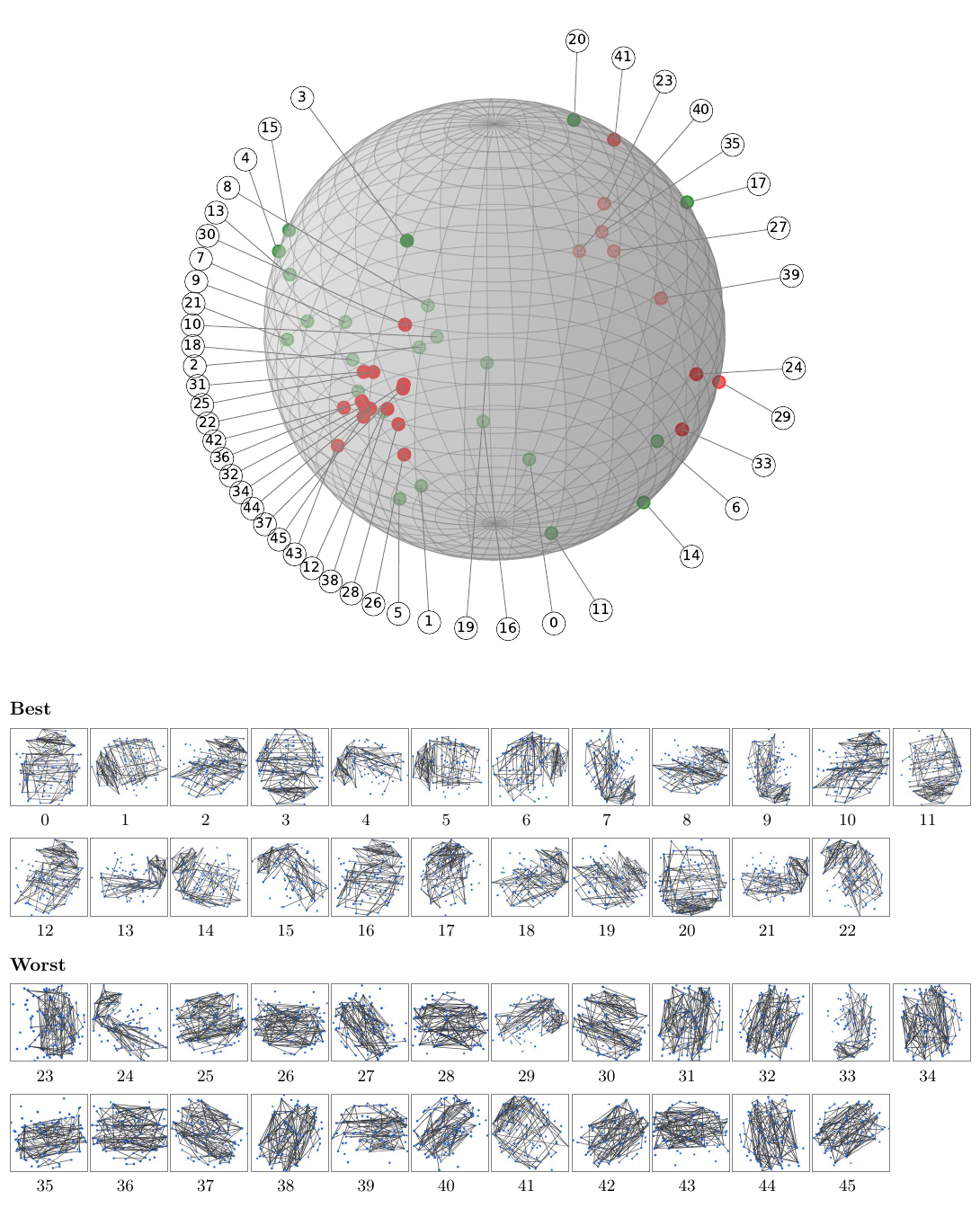}
    \caption{The distribution of all perspectives selected by users as \textit{best} (green) and \textit{worst} (red) mapped on a sphere surface for graph \textbf{S-7} (sem\_3\_100\_137) with $|V| = 100, \; |E| = 137$.}
\end{figure}

\begin{figure}
    \centering
    \includegraphics[width=1\linewidth]{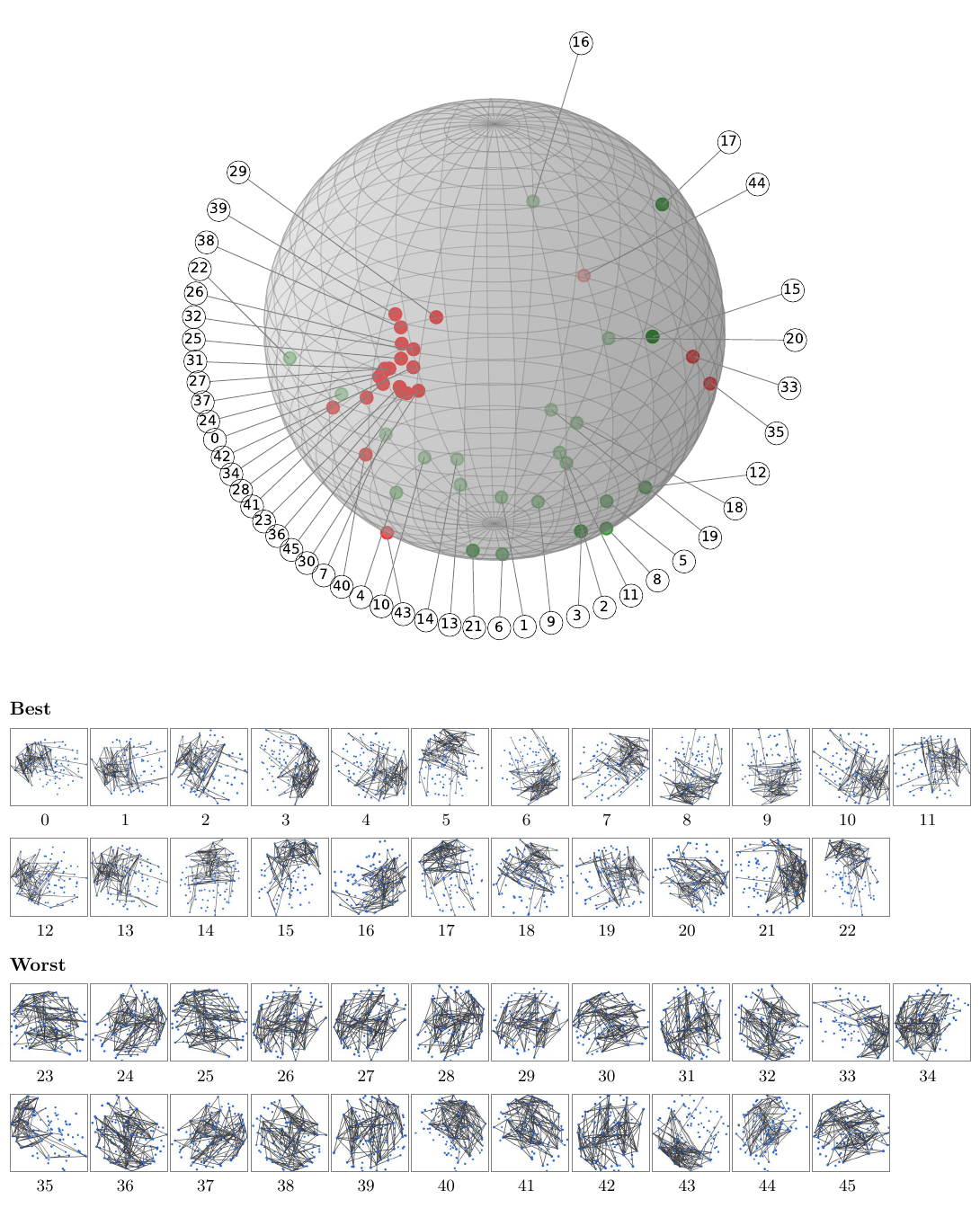}
    \caption{The distribution of all perspectives selected by users as \textit{best} (green) and \textit{worst} (red) mapped on a sphere surface for graph \textbf{S-8} (sem\_9\_100\_131) with $|V| = 100, \; |E| = 131$.}
\end{figure}

\begin{figure}
    \centering
    \includegraphics[width=1\linewidth]{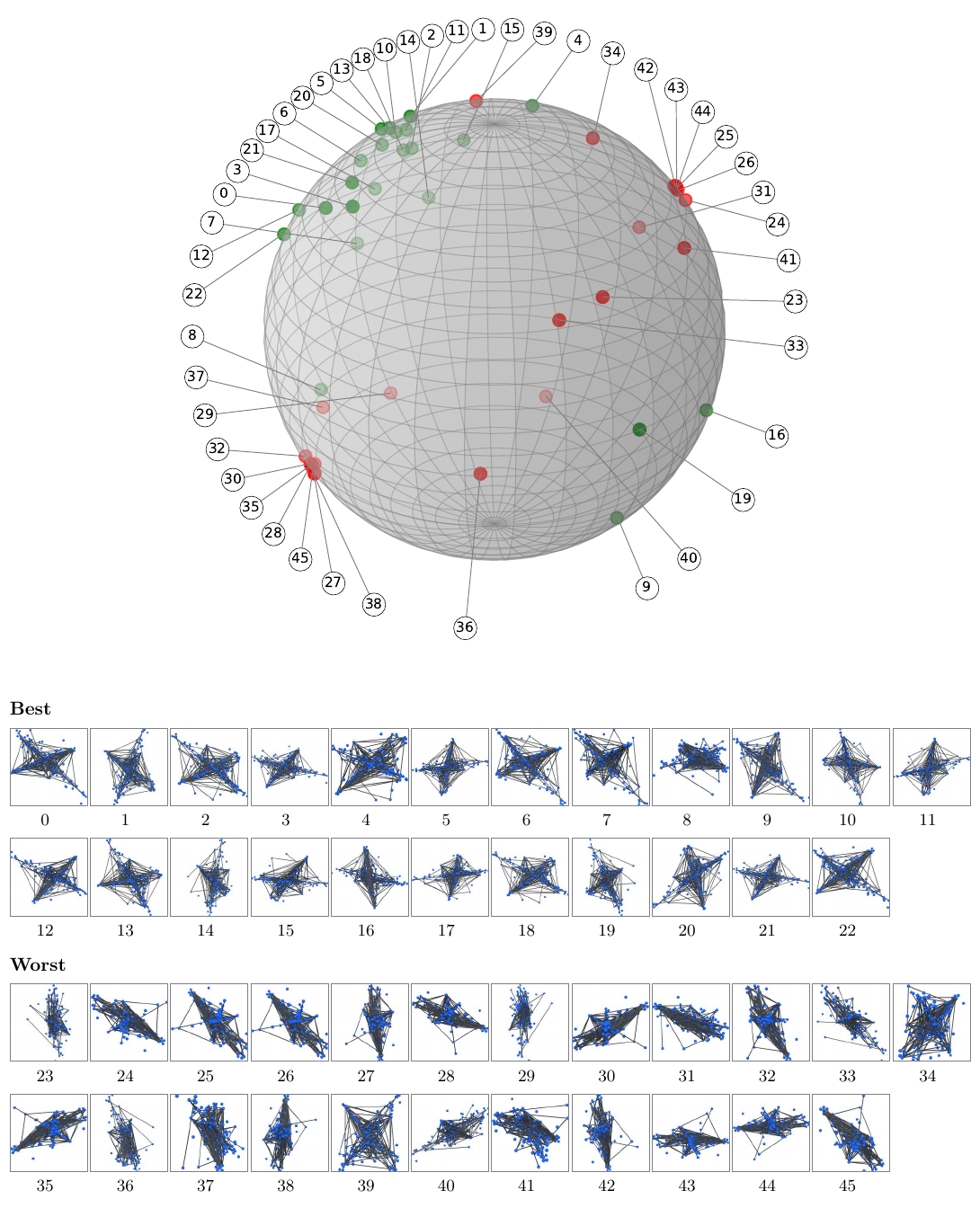}
    \caption{The distribution of all perspectives selected by users as \textit{best} (green) and \textit{worst} (red) mapped on a sphere surface for graph \textbf{S-9} (sem\_5\_200\_194) with $|V| = 200, \; |E| = 194$.}
\end{figure}

\begin{figure}
    \centering
    \includegraphics[width=1\linewidth]{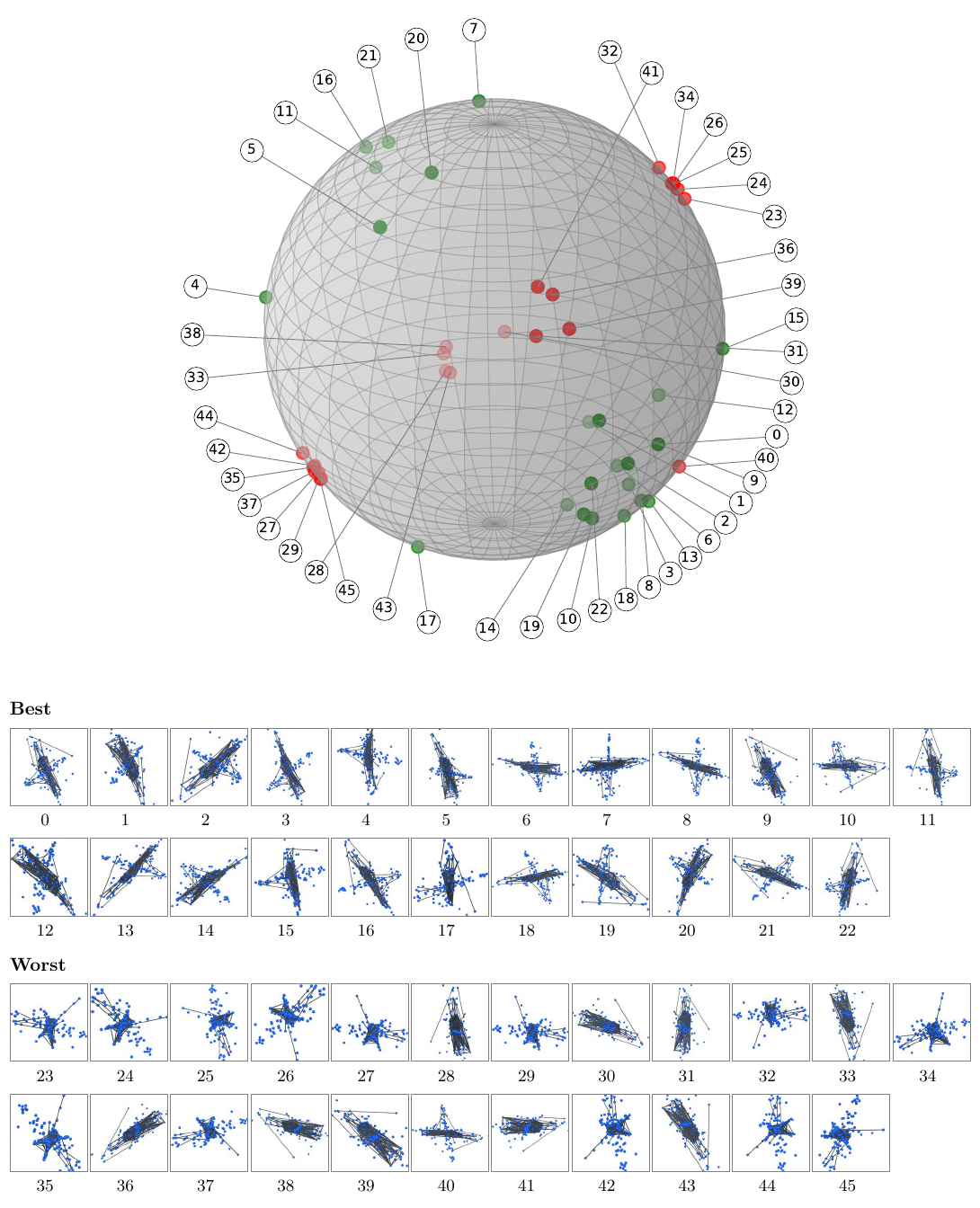}
    \caption{The distribution of all perspectives selected by users as \textit{best} (green) and \textit{worst} (red) mapped on a sphere surface for graph \textbf{S-10} (sem\_6\_200\_202) with $|V| = 200, \; |E| = 202$.}
\end{figure}

\begin{figure}
    \centering
    \includegraphics[width=1\linewidth]{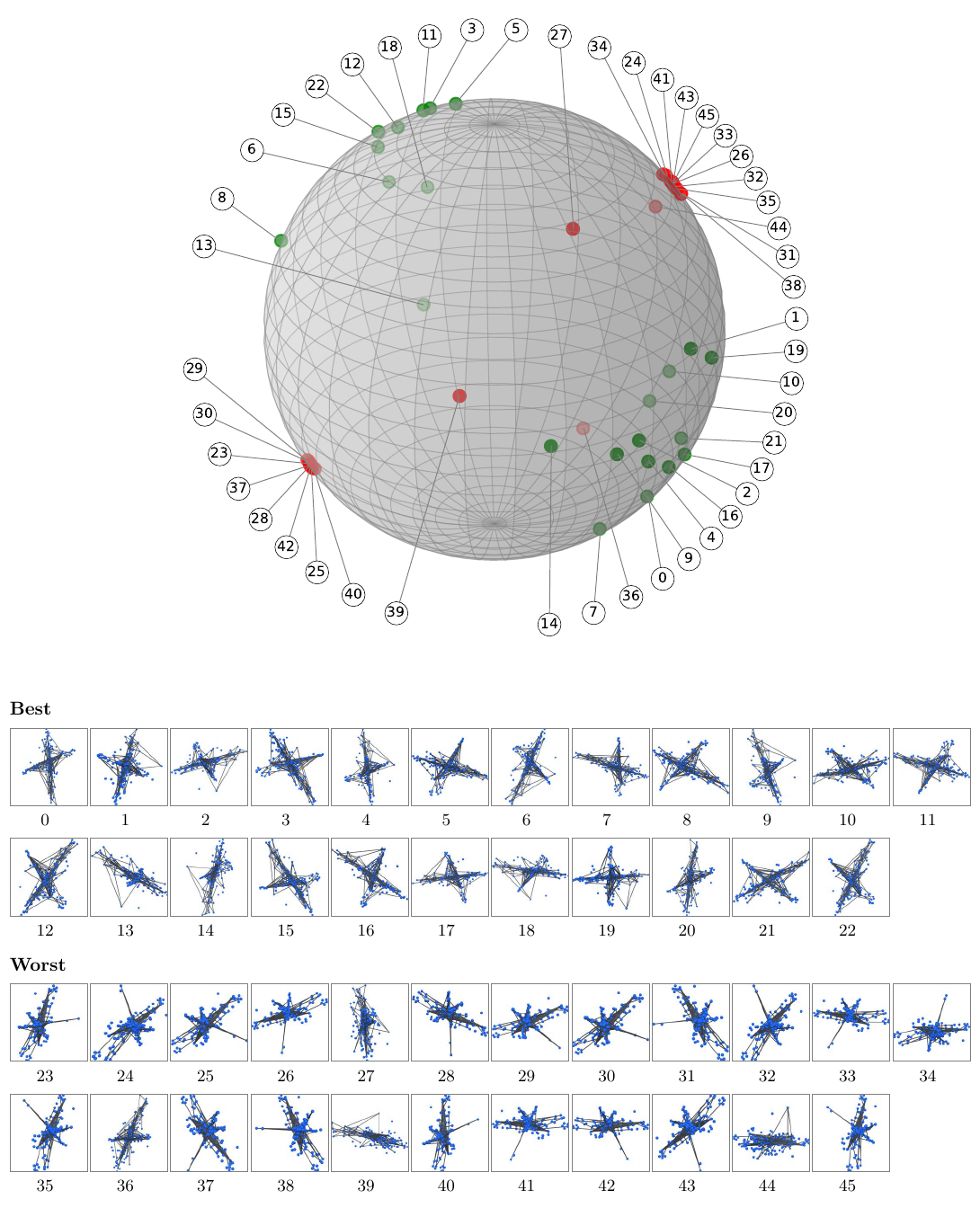}
    \caption{The distribution of all perspectives selected by users as \textit{best} (green) and \textit{worst} (red) mapped on a sphere surface for graph \textbf{S-11} (sem\_9\_200\_150) with $|V| = 200, \; |E| = 150$.}
\end{figure}

\begin{figure}
    \centering
    \includegraphics[width=1\linewidth]{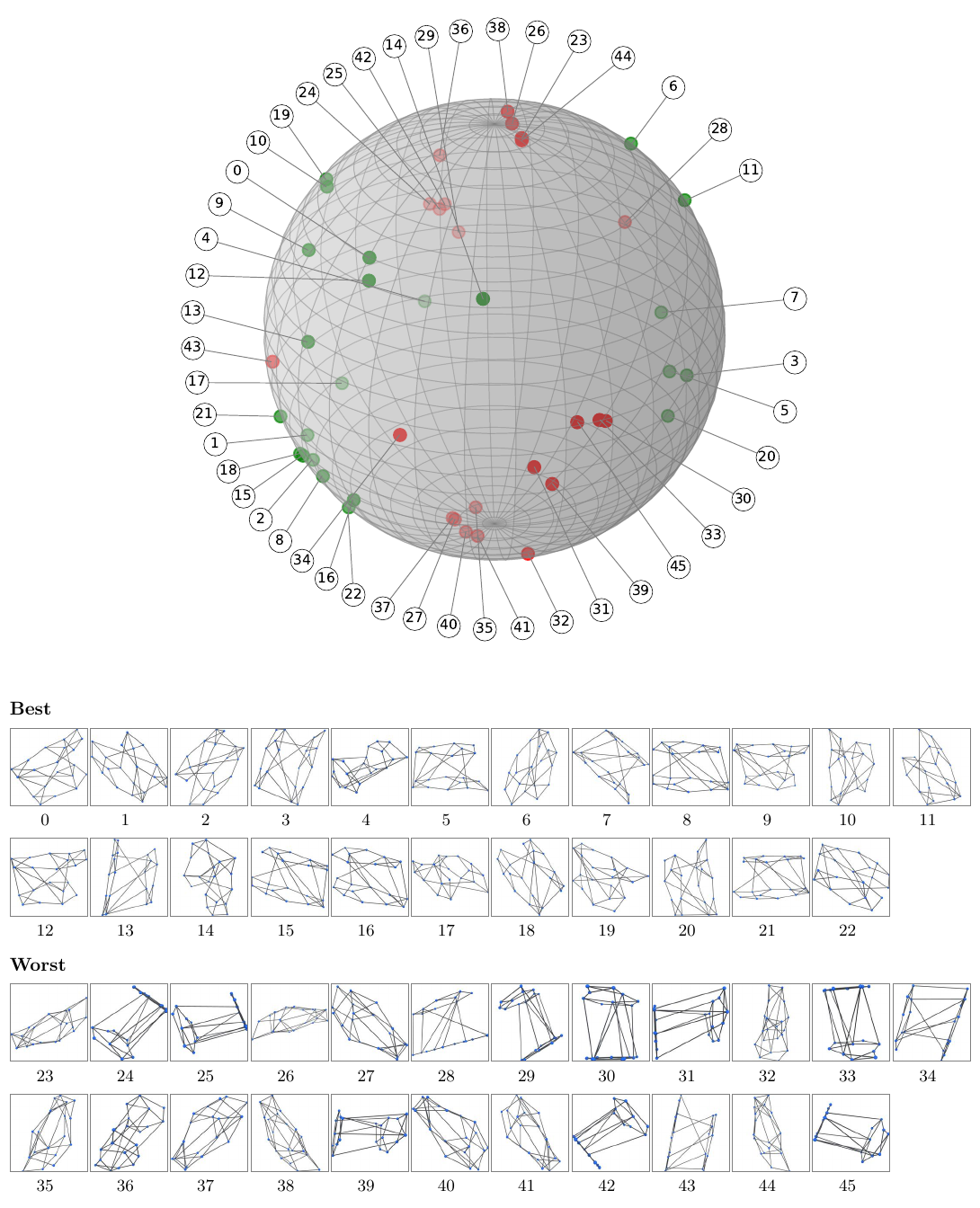}
    \caption{The distribution of all perspectives selected by users as \textit{best} (green) and \textit{worst} (red) mapped on a sphere surface for graph \textbf{L-0} (layered\_2\_20\_36) with $|V| = 20, \; |E| = 36$.}
\end{figure}

\begin{figure}
    \centering
    \includegraphics[width=1\linewidth]{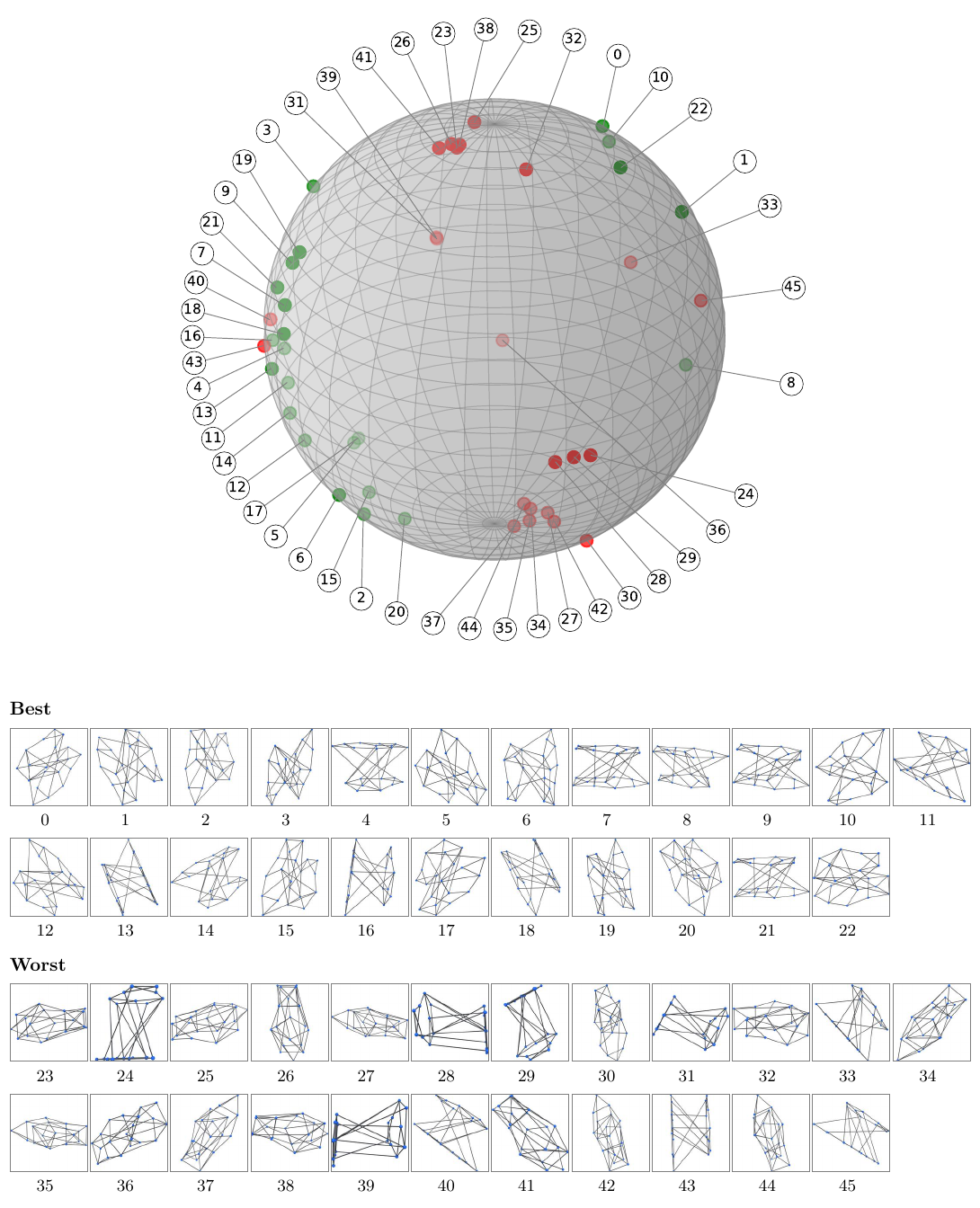}
    \caption{The distribution of all perspectives selected by users as \textit{best} (green) and \textit{worst} (red) mapped on a sphere surface for graph \textbf{L-1} (layered\_2\_20\_38) with $|V| = 20, \; |E| = 38$.}
\end{figure}

\begin{figure}
    \centering
    \includegraphics[width=1\linewidth]{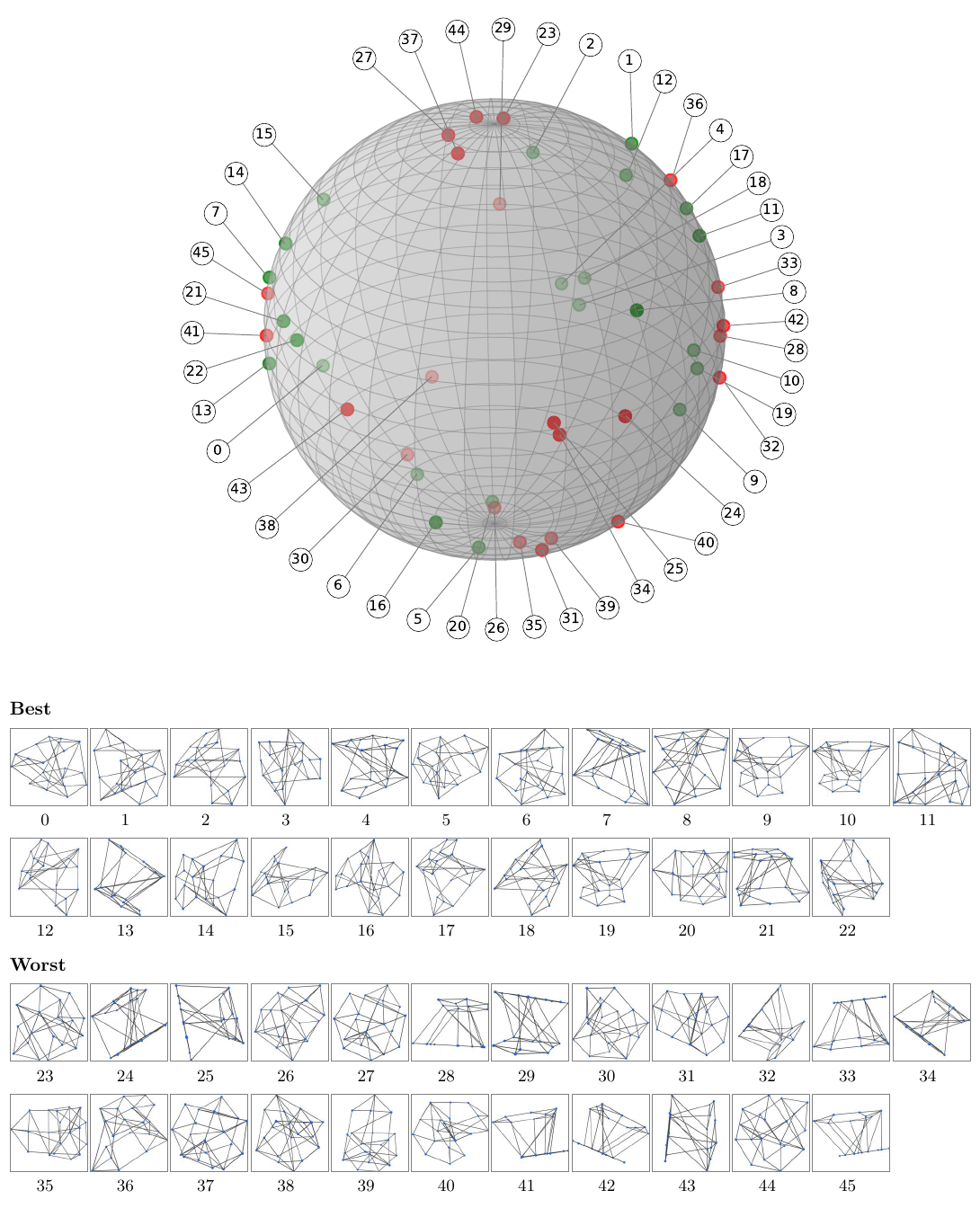}
    \caption{The distribution of all perspectives selected by users as \textit{best} (green) and \textit{worst} (red) mapped on a sphere surface for graph \textbf{L-2} (layered\_2\_20\_40) with $|V| = 20, \; |E| = 40$.}
\end{figure}

\begin{figure}
    \centering
    \includegraphics[width=1\linewidth]{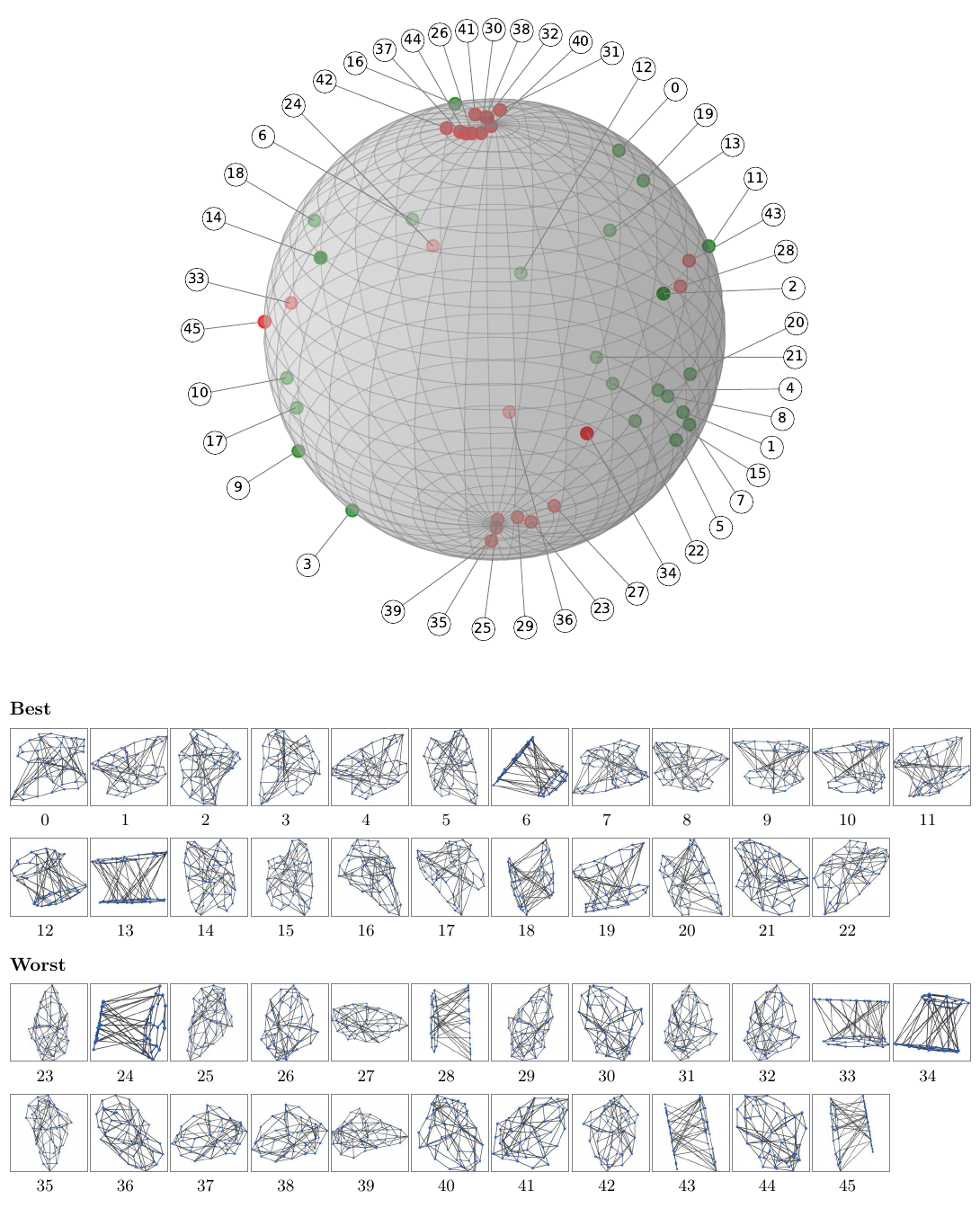}
    \caption{The distribution of all perspectives selected by users as \textit{best} (green) and \textit{worst} (red) mapped on a sphere surface for graph \textbf{L-3} (layered\_2\_50\_101) with $|V| = 50, \; |E| = 101$.}
\end{figure}

\begin{figure}
    \centering
    \includegraphics[width=1\linewidth]{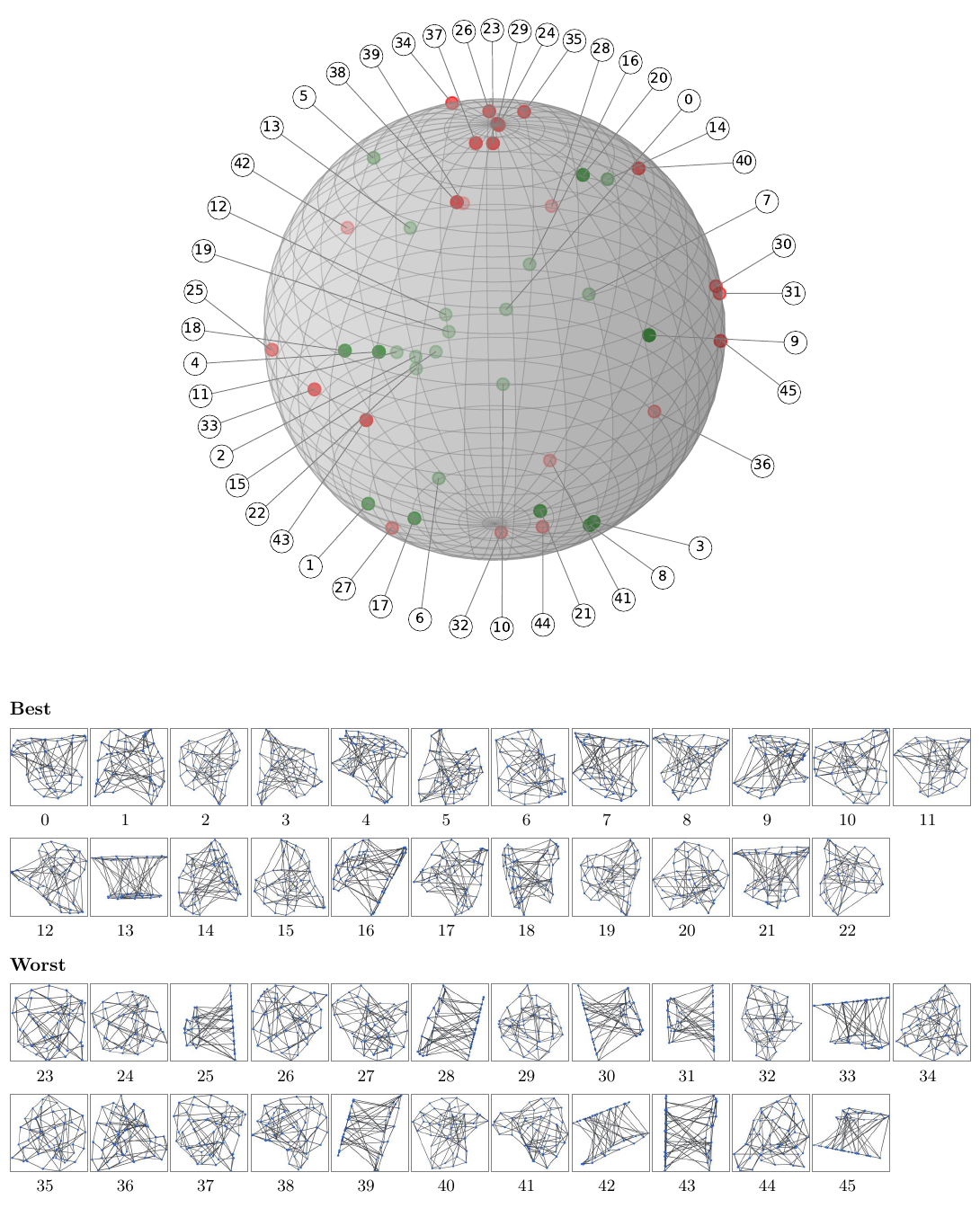}
    \caption{The distribution of all perspectives selected by users as \textit{best} (green) and \textit{worst} (red) mapped on a sphere surface for graph \textbf{L-4} (layered\_2\_50\_102) with $|V| = 50, \; |E| = 102$.}
\end{figure}

\begin{figure}
    \centering
    \includegraphics[width=1\linewidth]{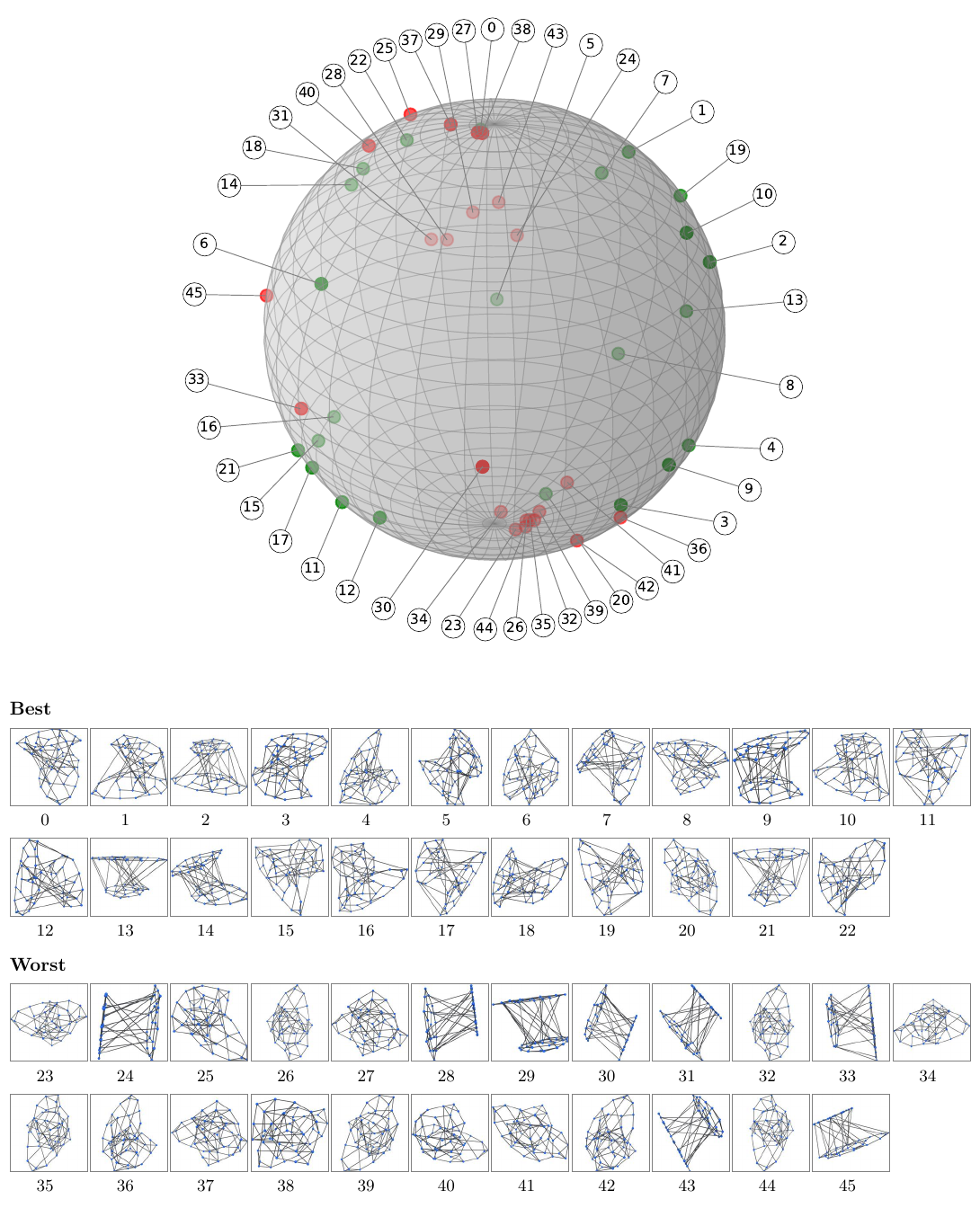}
    \caption{The distribution of all perspectives selected by users as \textit{best} (green) and \textit{worst} (red) mapped on a sphere surface for graph \textbf{L-5} (layered\_2\_50\_93) with $|V| = 50, \; |E| = 93$.}
\end{figure}

\begin{figure}
    \centering
    \includegraphics[width=1\linewidth]{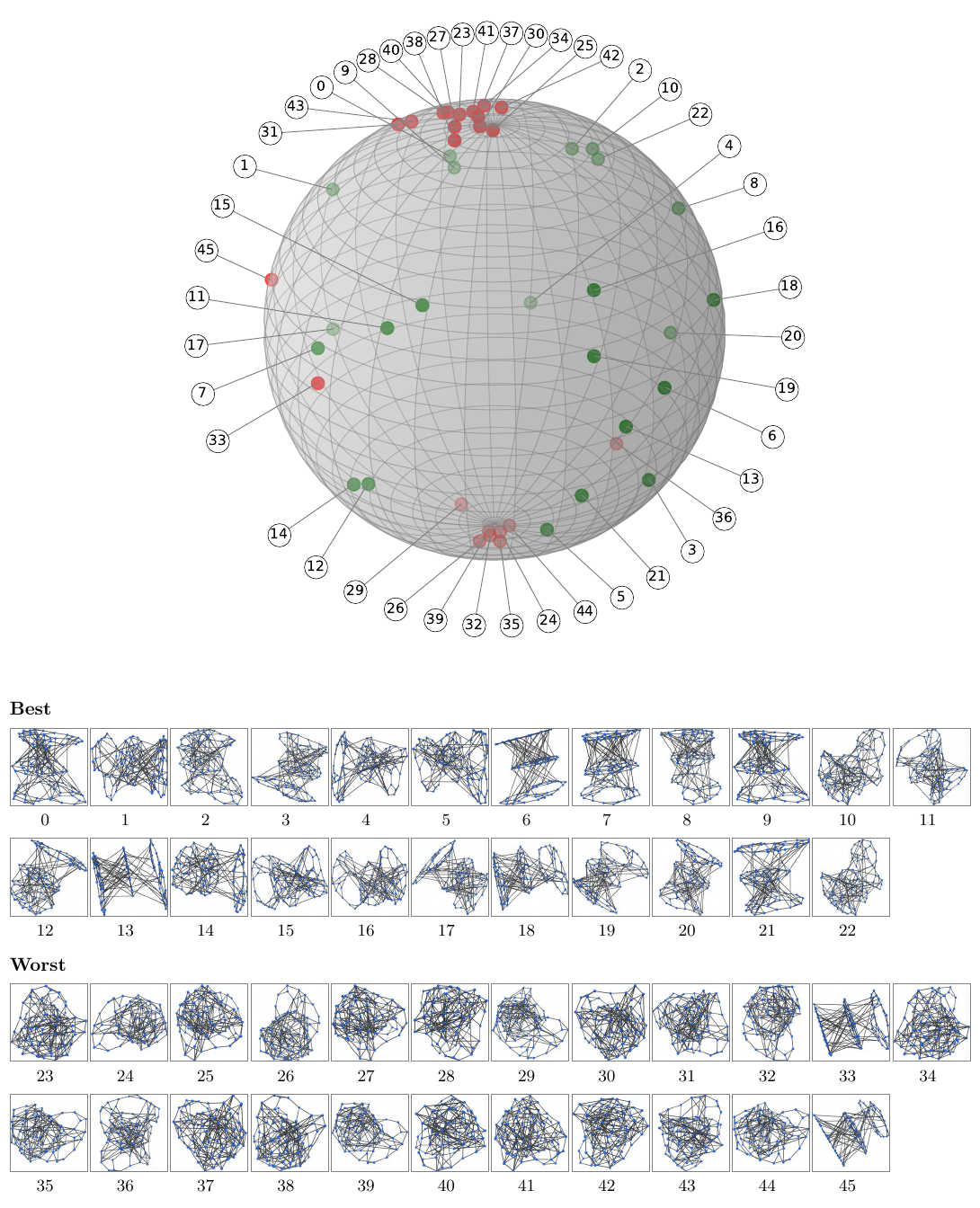}
    \caption{The distribution of all perspectives selected by users as \textit{best} (green) and \textit{worst} (red) mapped on a sphere surface for graph \textbf{L-6} (layered\_3\_100\_207) with $|V| = 100, \; |E| = 207$.}
\end{figure}

\begin{figure}
    \centering
    \includegraphics[width=1\linewidth]{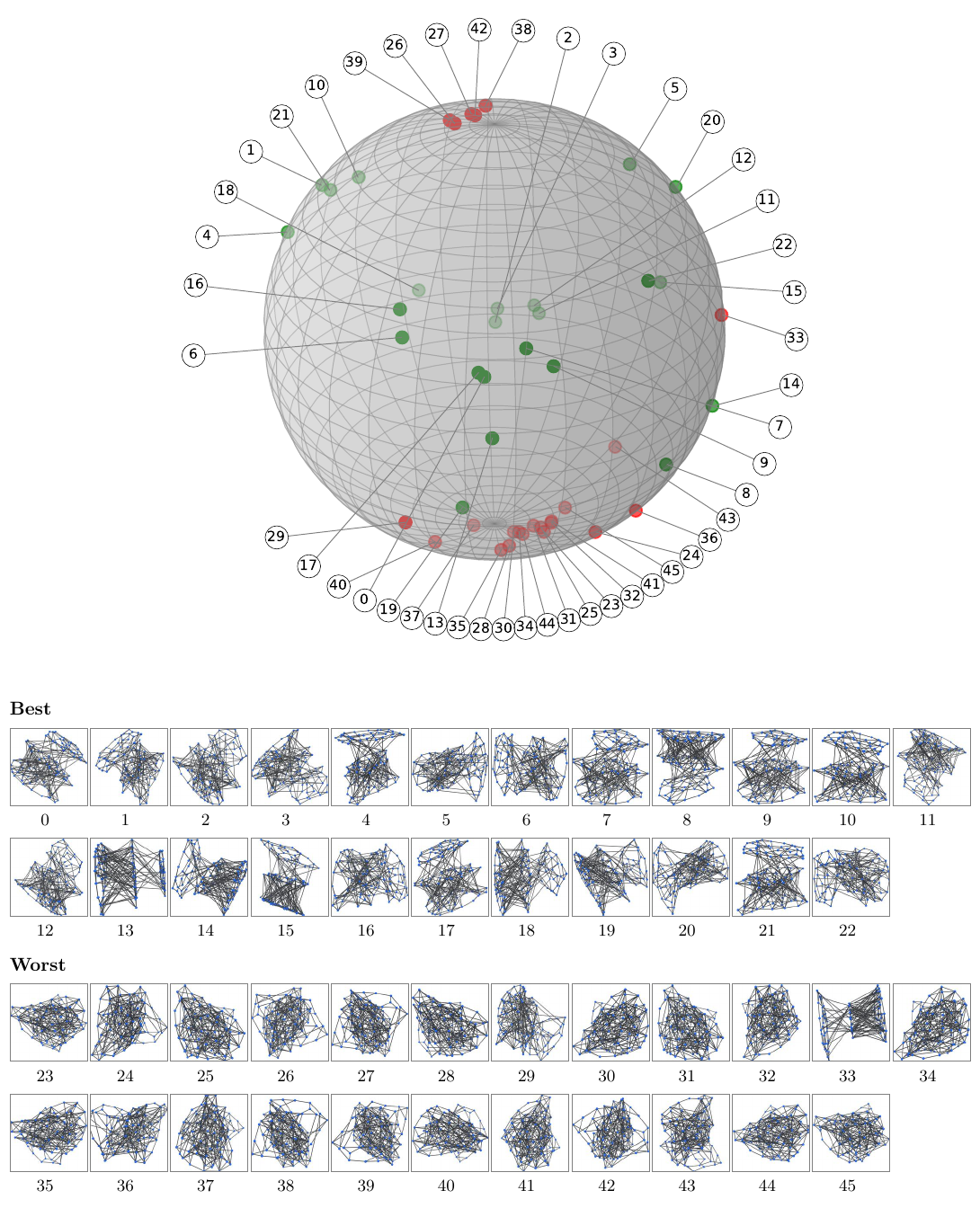}
    \caption{The distribution of all perspectives selected by users as \textit{best} (green) and \textit{worst} (red) mapped on a sphere surface for graph \textbf{L-7} (layered\_3\_100\_253) with $|V| = 100, \; |E| = 253$.}
\end{figure}

\begin{figure}
    \centering
    \includegraphics[width=1\linewidth]{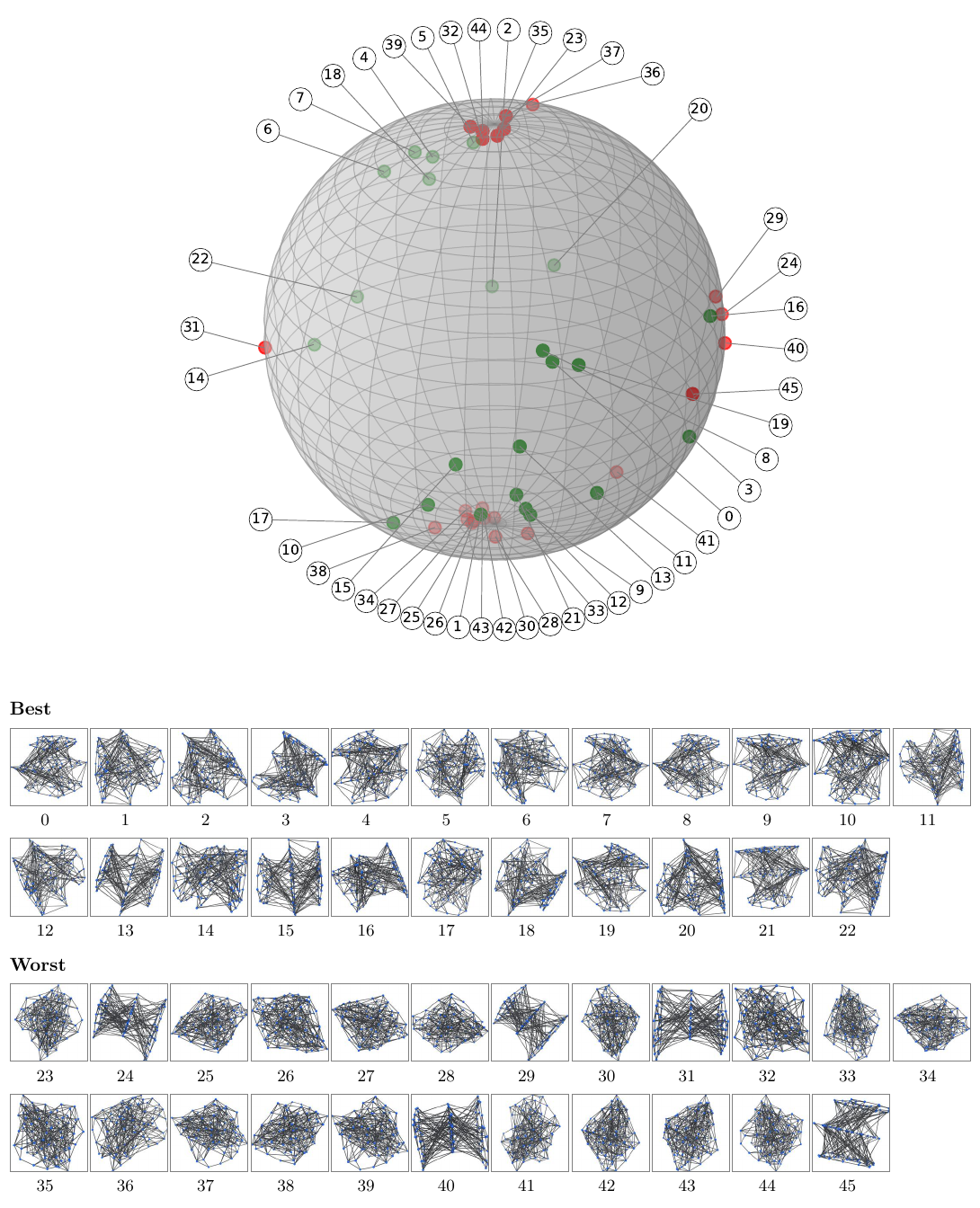}
    \caption{The distribution of all perspectives selected by users as \textit{best} (green) and \textit{worst} (red) mapped on a sphere surface for graph \textbf{L-8} (layered\_3\_100\_284) with $|V| = 100, \; |E| = 284$.}
\end{figure}

\begin{figure}
    \centering
    \includegraphics[width=1\linewidth]{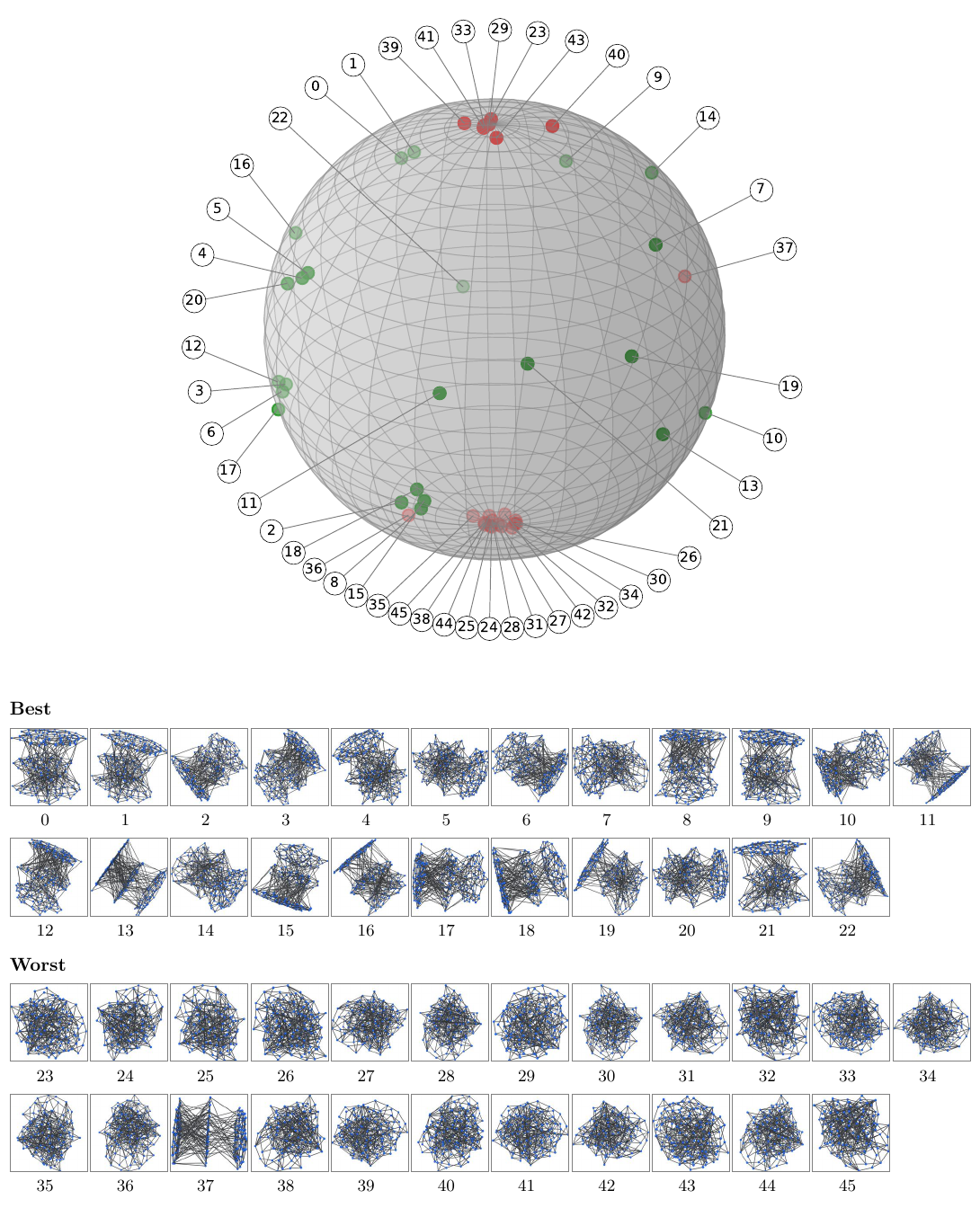}
    \caption{The distribution of all perspectives selected by users as \textit{best} (green) and \textit{worst} (red) mapped on a sphere surface for graph \textbf{L-9} (layered\_3\_200\_431) with $|V| = 200, \; |E| = 431$.}
\end{figure}

\begin{figure}
    \centering
    \includegraphics[width=1\linewidth]{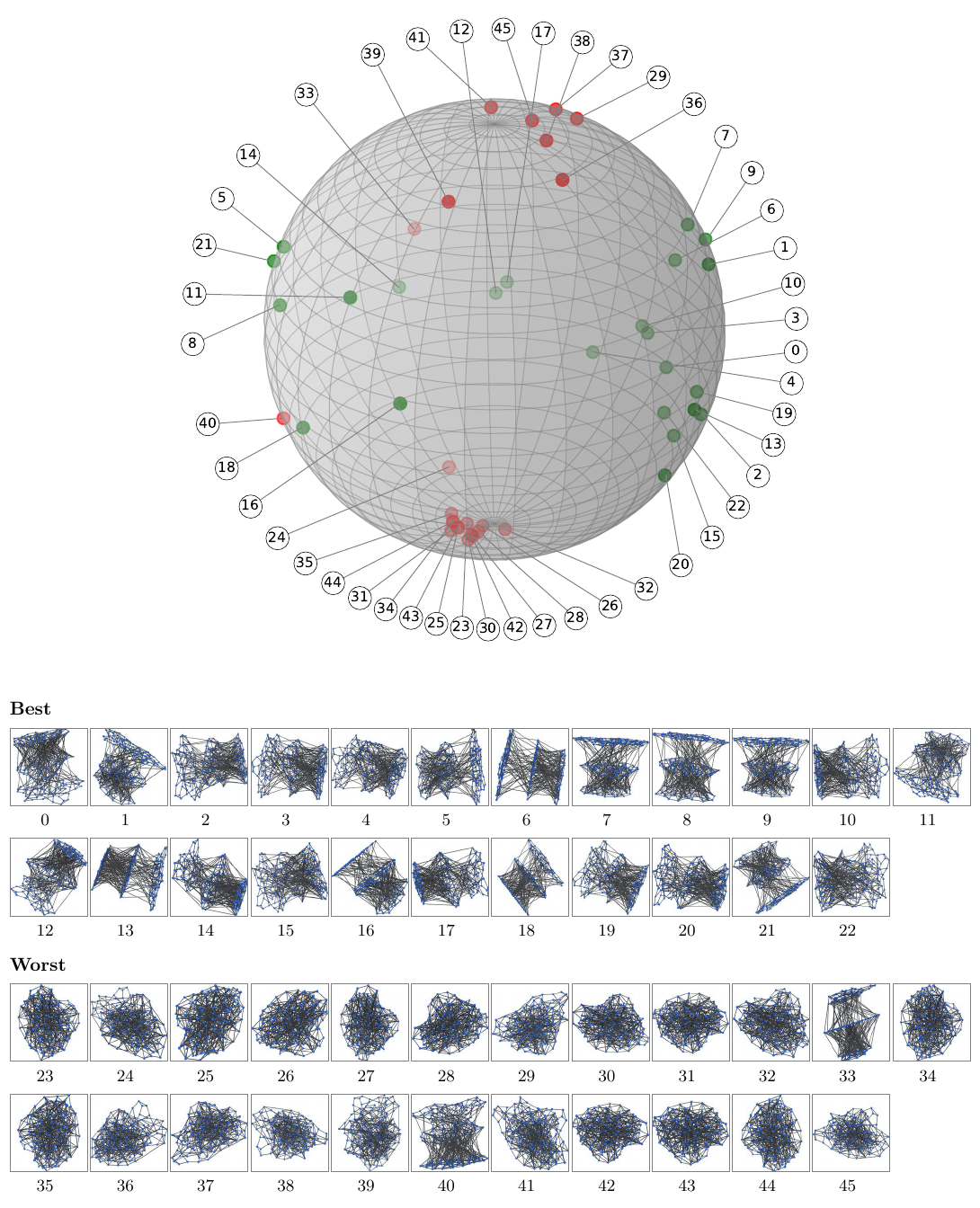}
    \caption{The distribution of all perspectives selected by users as \textit{best} (green) and \textit{worst} (red) mapped on a sphere surface for graph \textbf{L-10} (layered\_3\_200\_487) with $|V| = 200, \; |E| = 487$.}
\end{figure}

\begin{figure}
    \centering
    \includegraphics[width=1\linewidth]{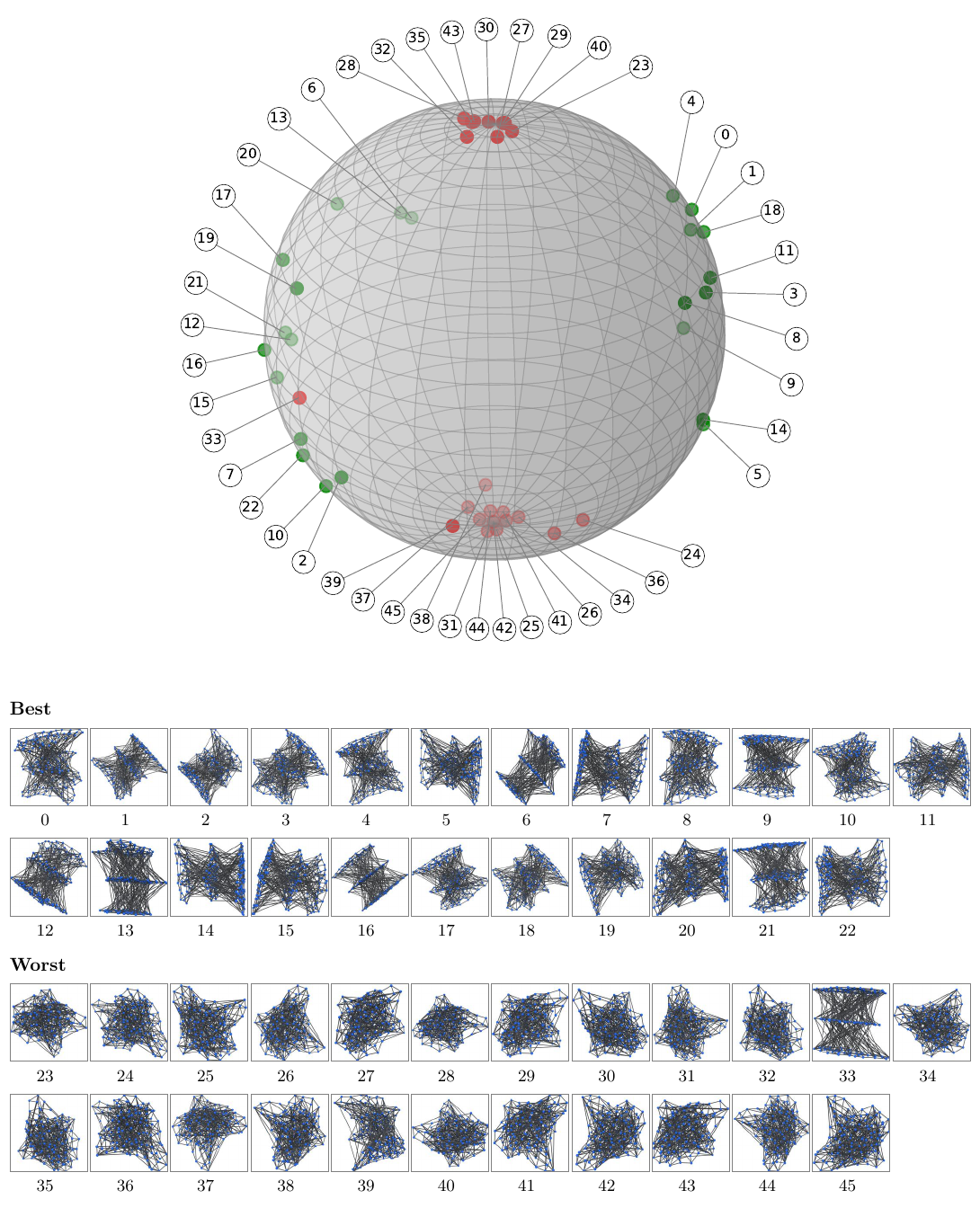}
    \caption{The distribution of all perspectives selected by users as \textit{best} (green) and \textit{worst} (red) mapped on a sphere surface for graph \textbf{L-11} (layered\_3\_200\_508) with $|V| = 200, \; |E| = 508$.}
\end{figure}

\begin{figure}
    \centering
    \includegraphics[width=1\linewidth]{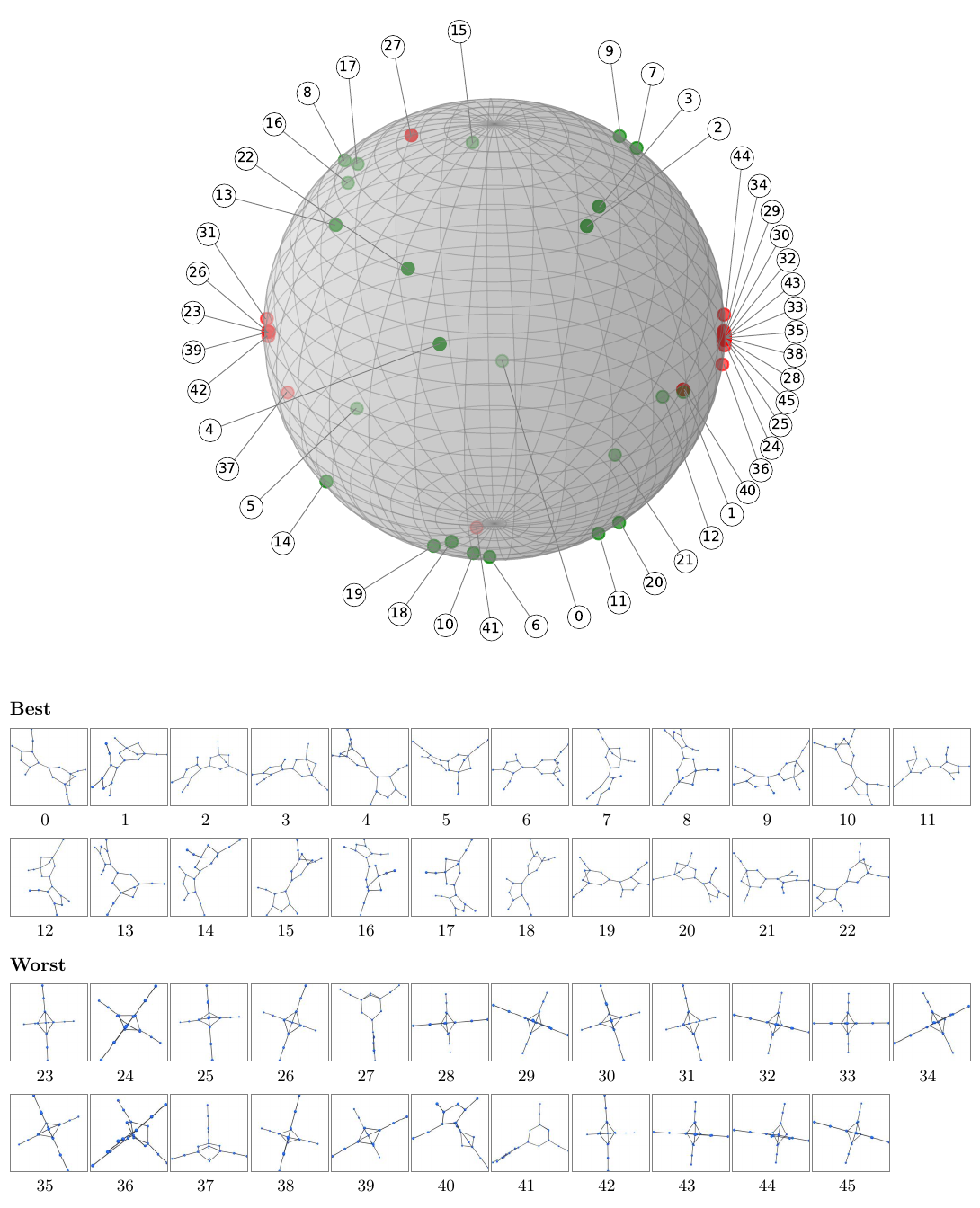}
    \caption{The distribution of all perspectives selected by users as \textit{best} (green) and \textit{worst} (red) mapped on a sphere surface for graph \textbf{E-0} (grafo2347.20) with $|V| = 20, \; |E| = 22$.}
\end{figure}

\begin{figure}
    \centering
    \includegraphics[width=1\linewidth]{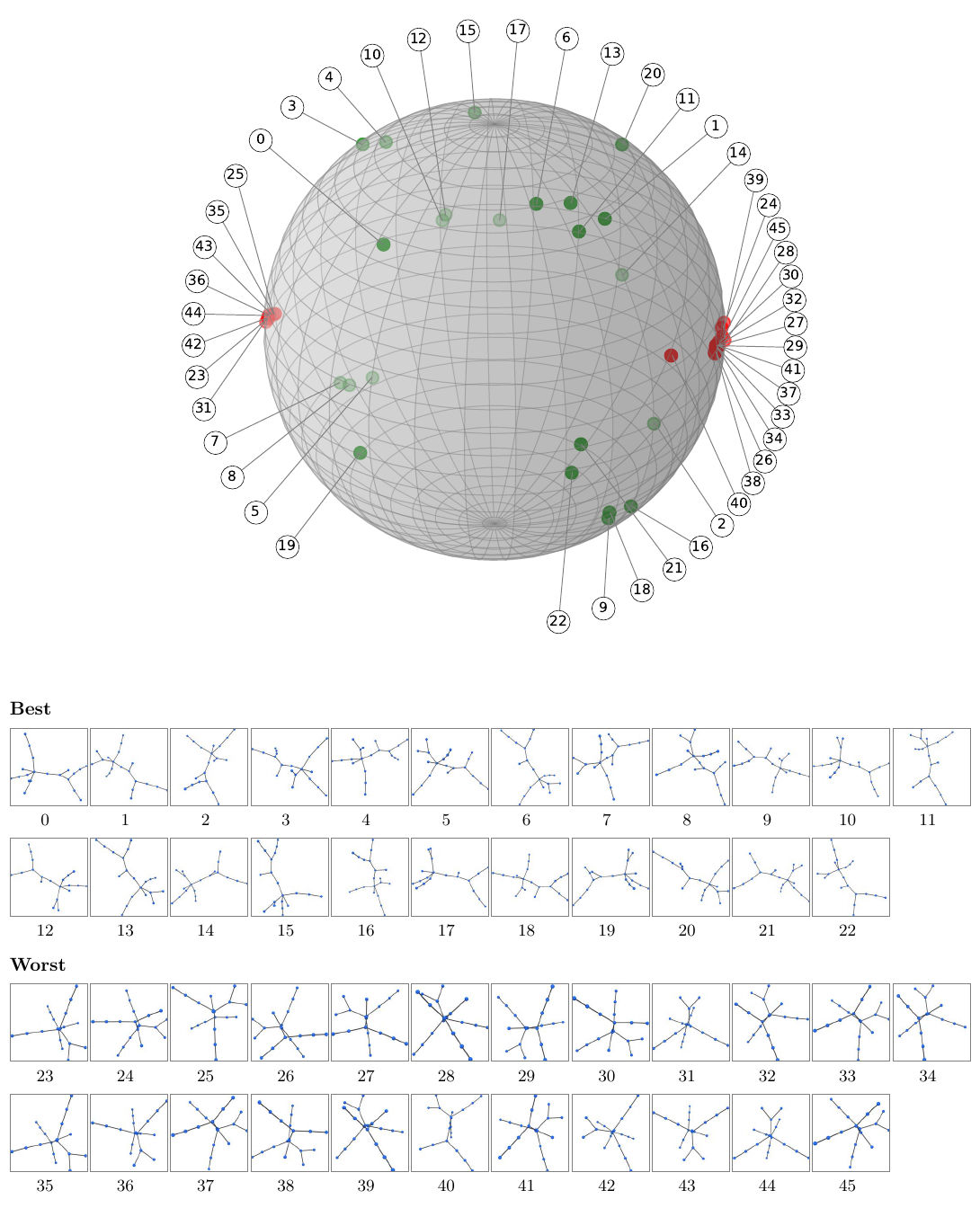}
    \caption{The distribution of all perspectives selected by users as \textit{best} (green) and \textit{worst} (red) mapped on a sphere surface for graph \textbf{E-1} (tree\_20\_0) with $|V| = 20, \; |E| = 19$.}
\end{figure}

\begin{figure}
    \centering
    \includegraphics[width=1\linewidth]{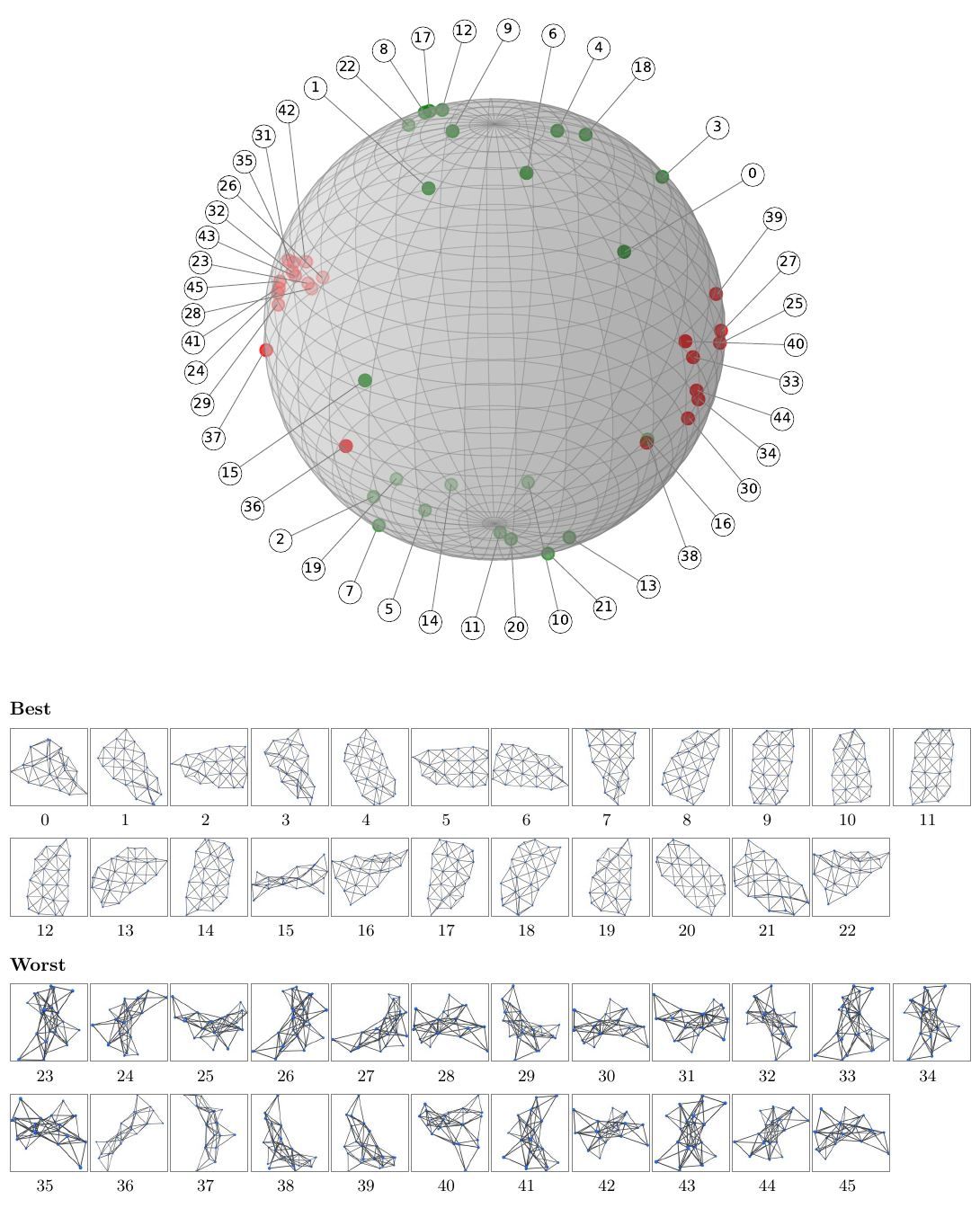}
    \caption{The distribution of all perspectives selected by users as \textit{best} (green) and \textit{worst} (red) mapped on a sphere surface for graph \textbf{E-2} (can\_24) with $|V| = 24, \; |E| = 92$.}
\end{figure}

\begin{figure}
    \centering
    \includegraphics[width=1\linewidth]{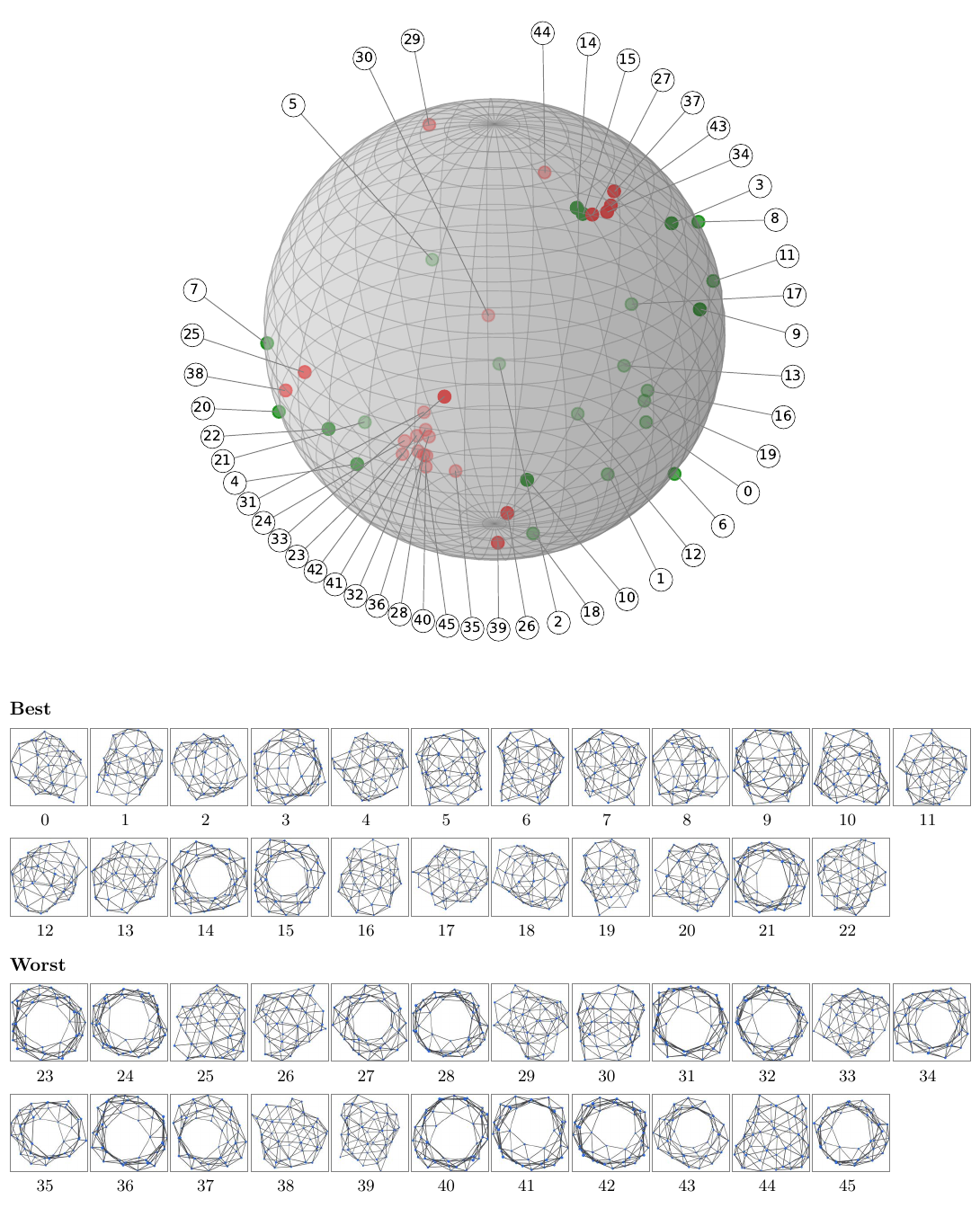}
    \caption{The distribution of all perspectives selected by users as \textit{best} (green) and \textit{worst} (red) mapped on a sphere surface for graph \textbf{E-3} (mesh1em1) with $|V| = 48, \; |E| = 177$.}
\end{figure}

\begin{figure}
    \centering
    \includegraphics[width=1\linewidth]{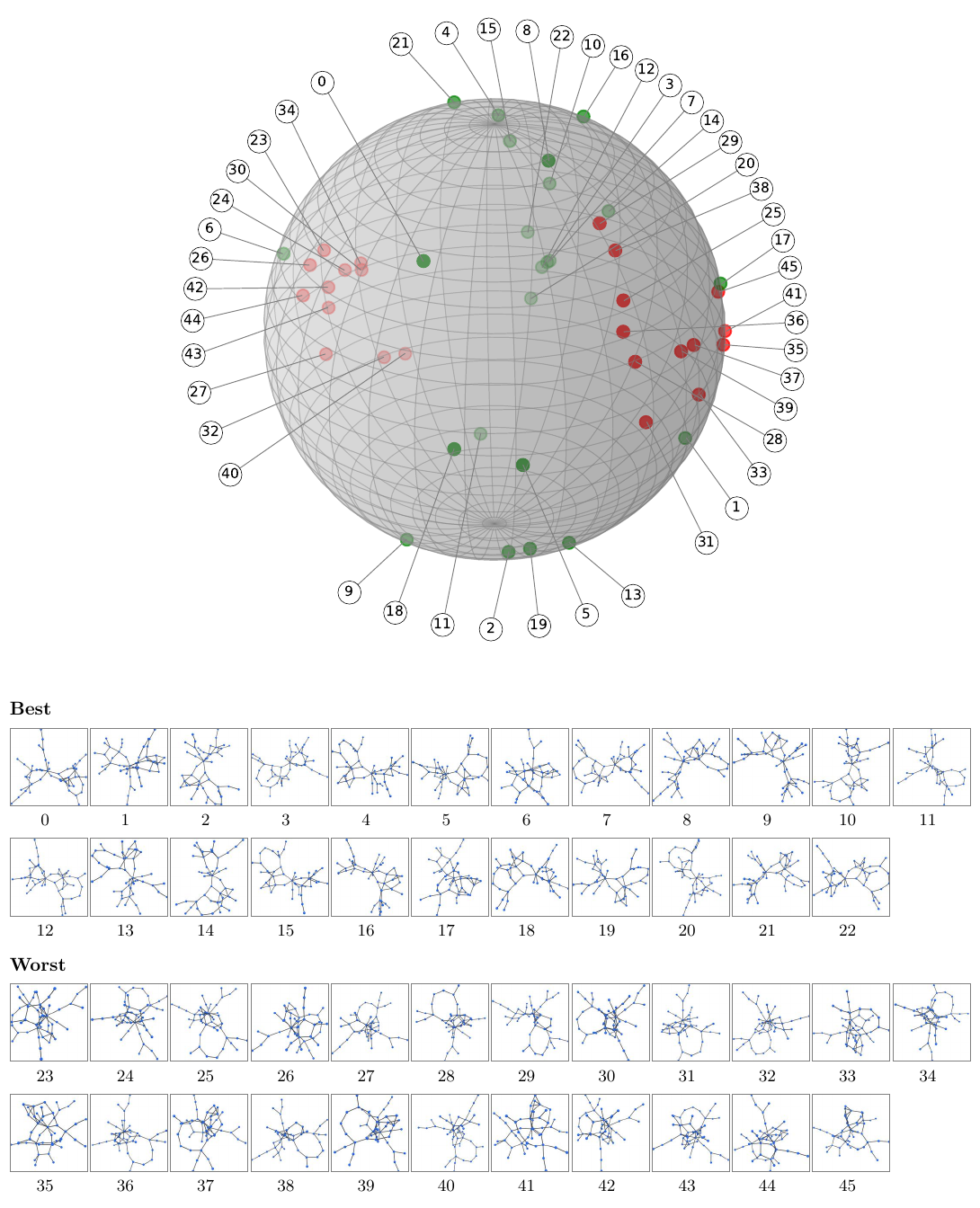}
    \caption{The distribution of all perspectives selected by users as \textit{best} (green) and \textit{worst} (red) mapped on a sphere surface for graph \textbf{E-4} (grafo5640.50) with $|V| = 50, \; |E| = 58$.}
\end{figure}

\begin{figure}
    \centering
    \includegraphics[width=1\linewidth]{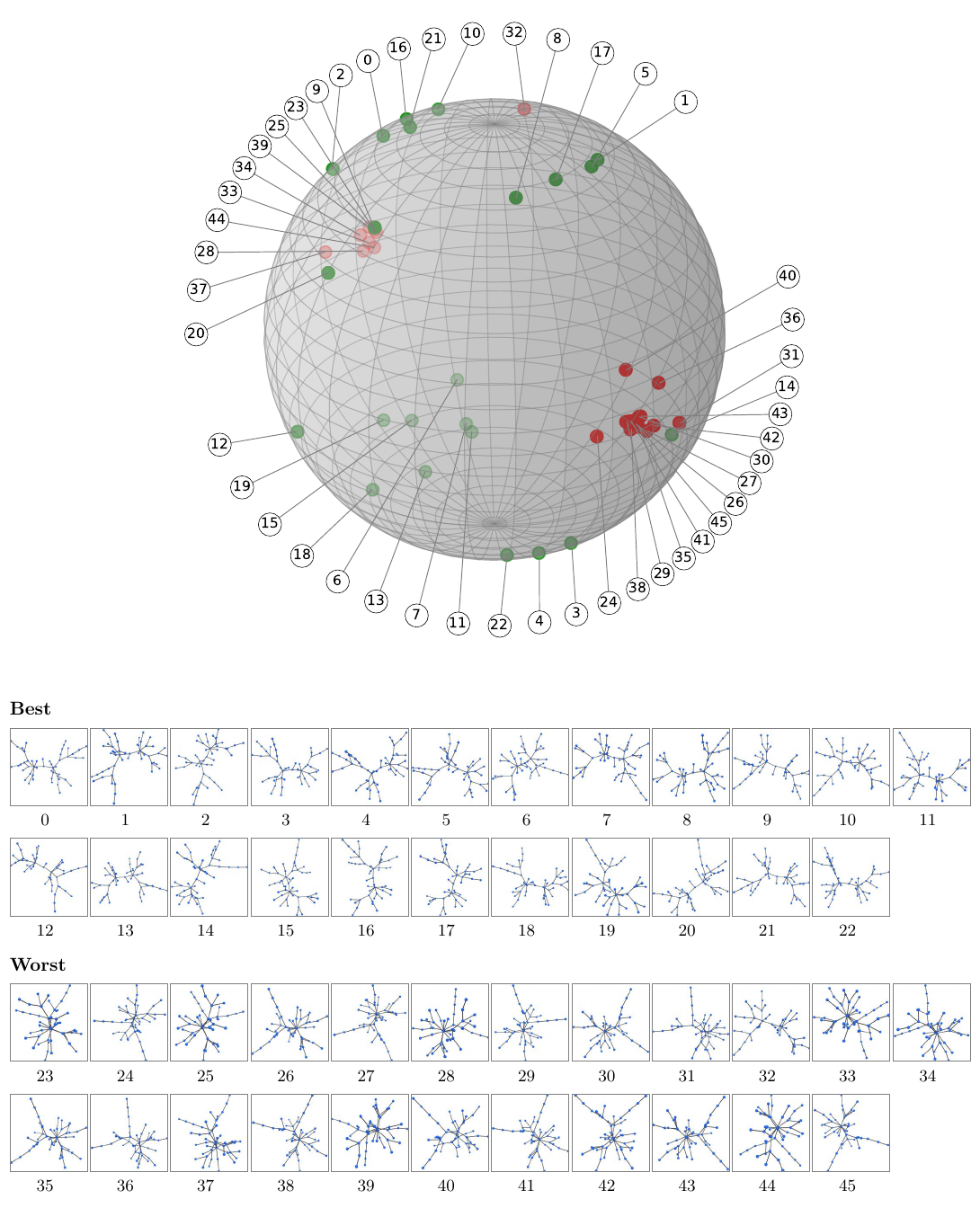}
    \caption{The distribution of all perspectives selected by users as \textit{best} (green) and \textit{worst} (red) mapped on a sphere surface for graph \textbf{E-5} (tree\_50\_2) with $|V| = 50, \; |E| = 49$.}
\end{figure}

\begin{figure}
    \centering
    \includegraphics[width=1\linewidth]{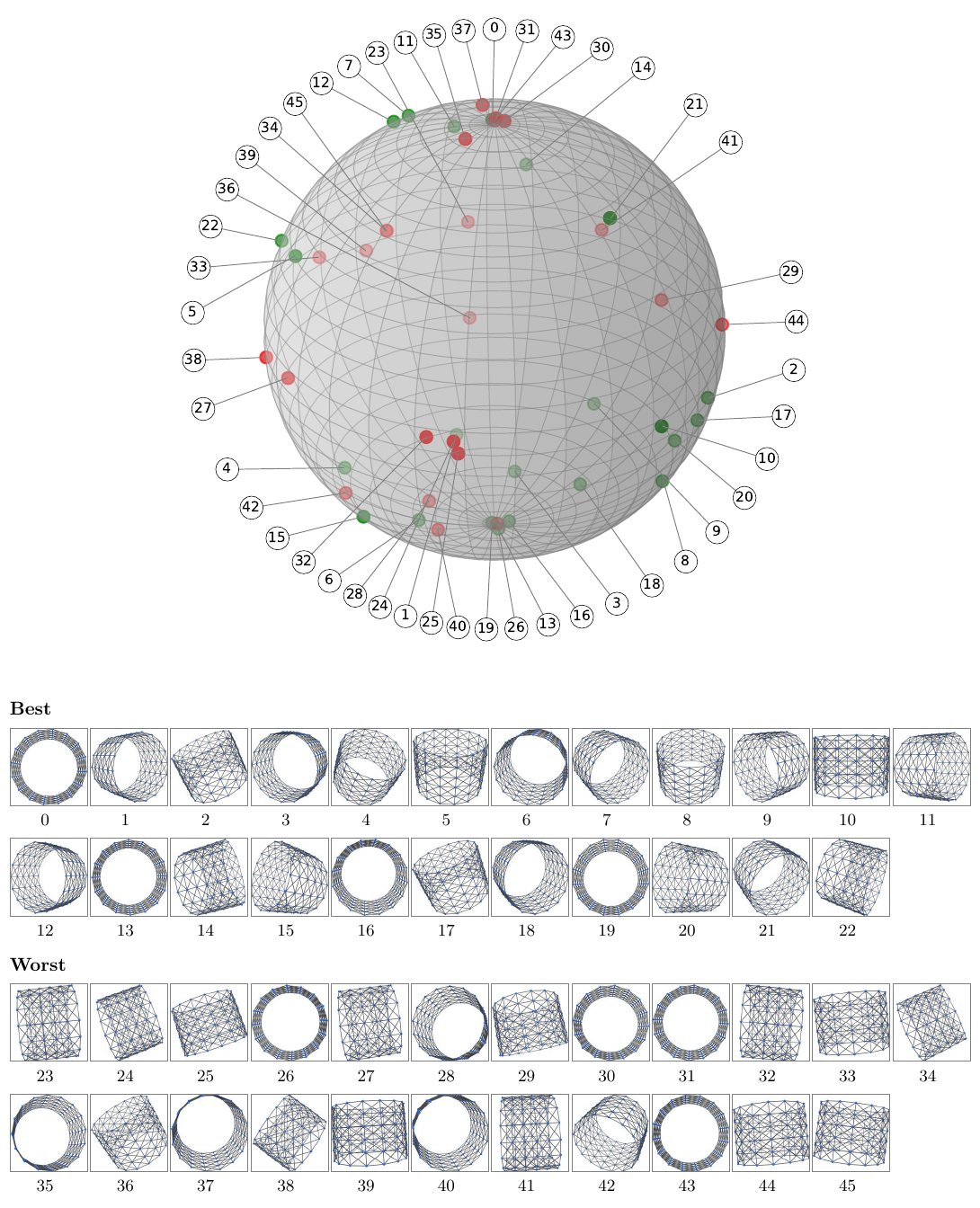}
    \caption{The distribution of all perspectives selected by users as \textit{best} (green) and \textit{worst} (red) mapped on a sphere surface for graph \textbf{E-6} (can\_96) with $|V| = 96, \; |E| = 432$.}
\end{figure}

\begin{figure}
    \centering
    \includegraphics[width=1\linewidth]{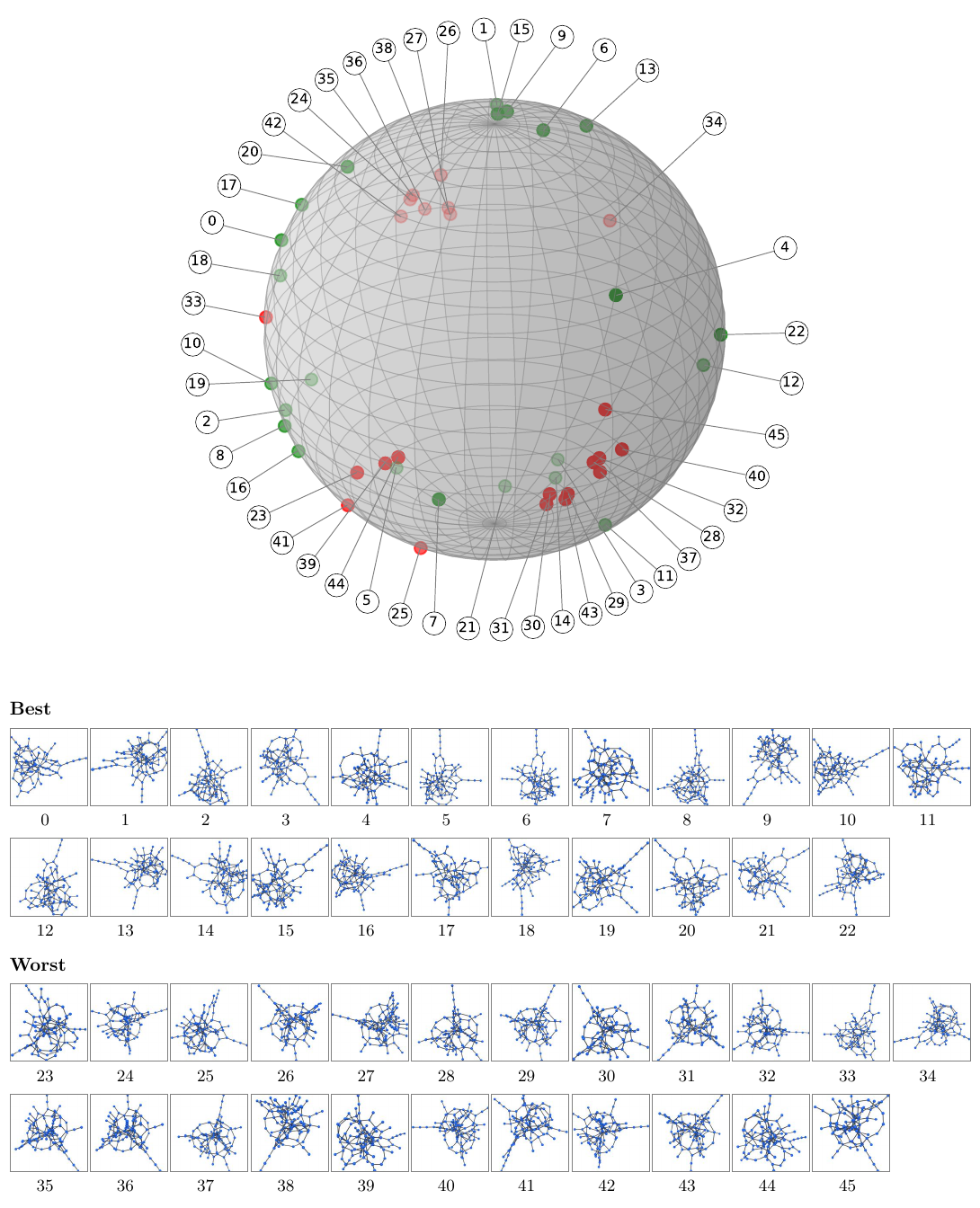}
    \caption{The distribution of all perspectives selected by users as \textit{best} (green) and \textit{worst} (red) mapped on a sphere surface for graph \textbf{E-7} (grafo10183.100) with $|V| = 100, \; |E| = 132$.}
\end{figure}

\begin{figure}
    \centering
    \includegraphics[width=1\linewidth]{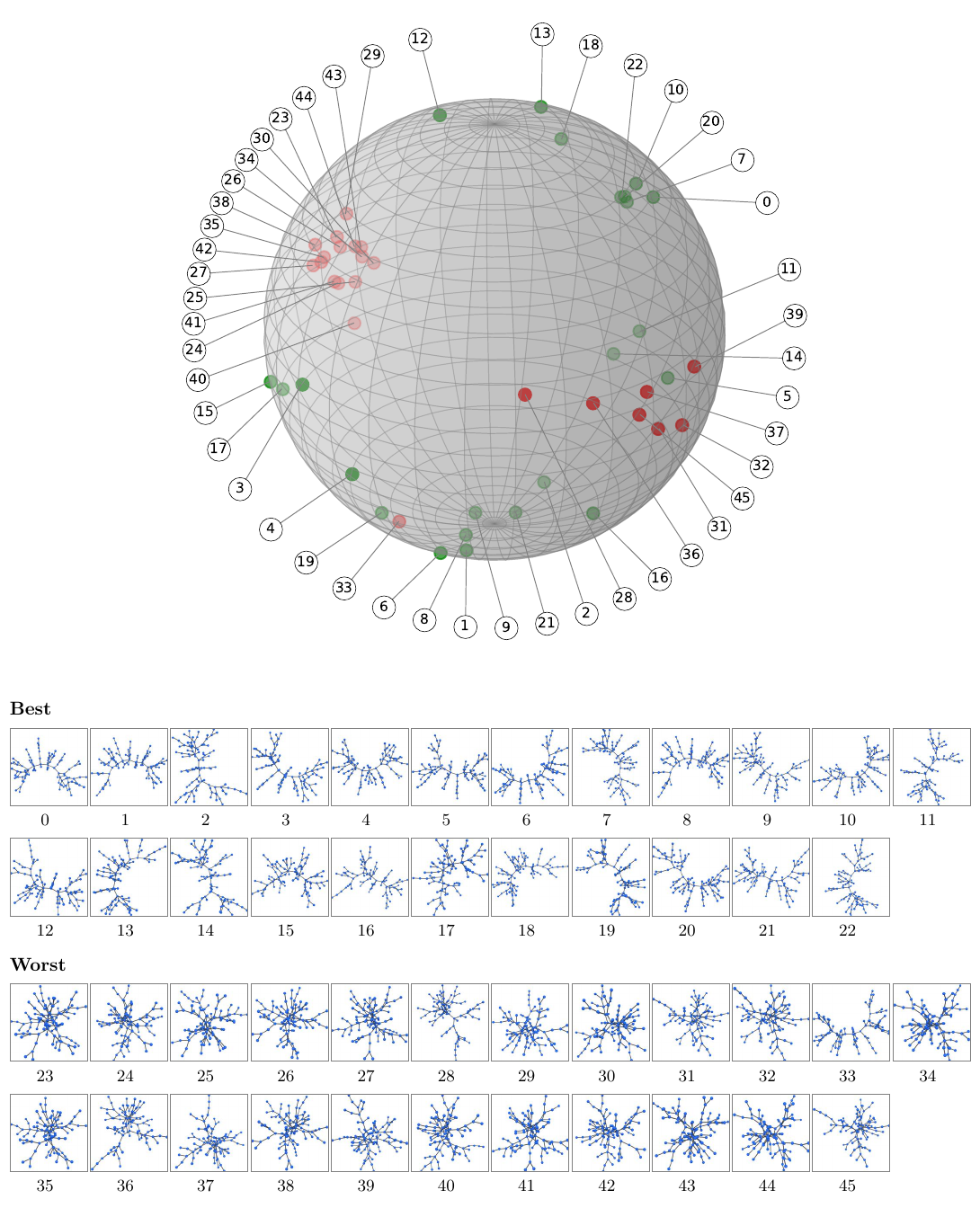}
    \caption{The distribution of all perspectives selected by users as \textit{best} (green) and \textit{worst} (red) mapped on a sphere surface for graph \textbf{E-8} (tree\_100\_2) with $|V| = 100, \; |E| = 99$.}
\end{figure}

\begin{figure}
    \centering
    \includegraphics[width=1\linewidth]{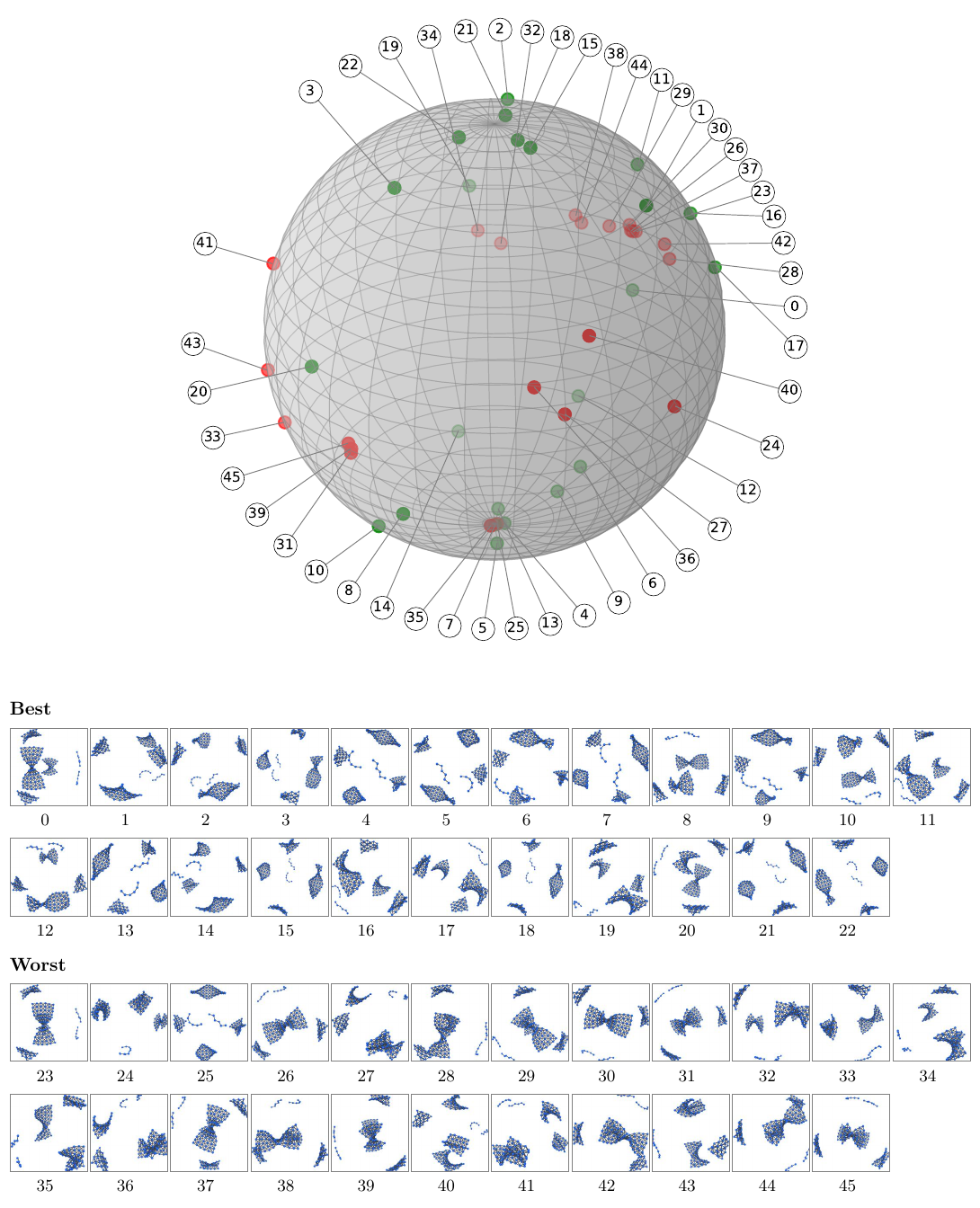}
    \caption{The distribution of all perspectives selected by users as \textit{best} (green) and \textit{worst} (red) mapped on a sphere surface for graph \textbf{E-9} (dwt\_198) with $|V| = 198, \; |E| = 795$.}
    \label{fig:appendix-b-e-9}
\end{figure}

\begin{figure}
    \centering
    \includegraphics[width=1\linewidth]{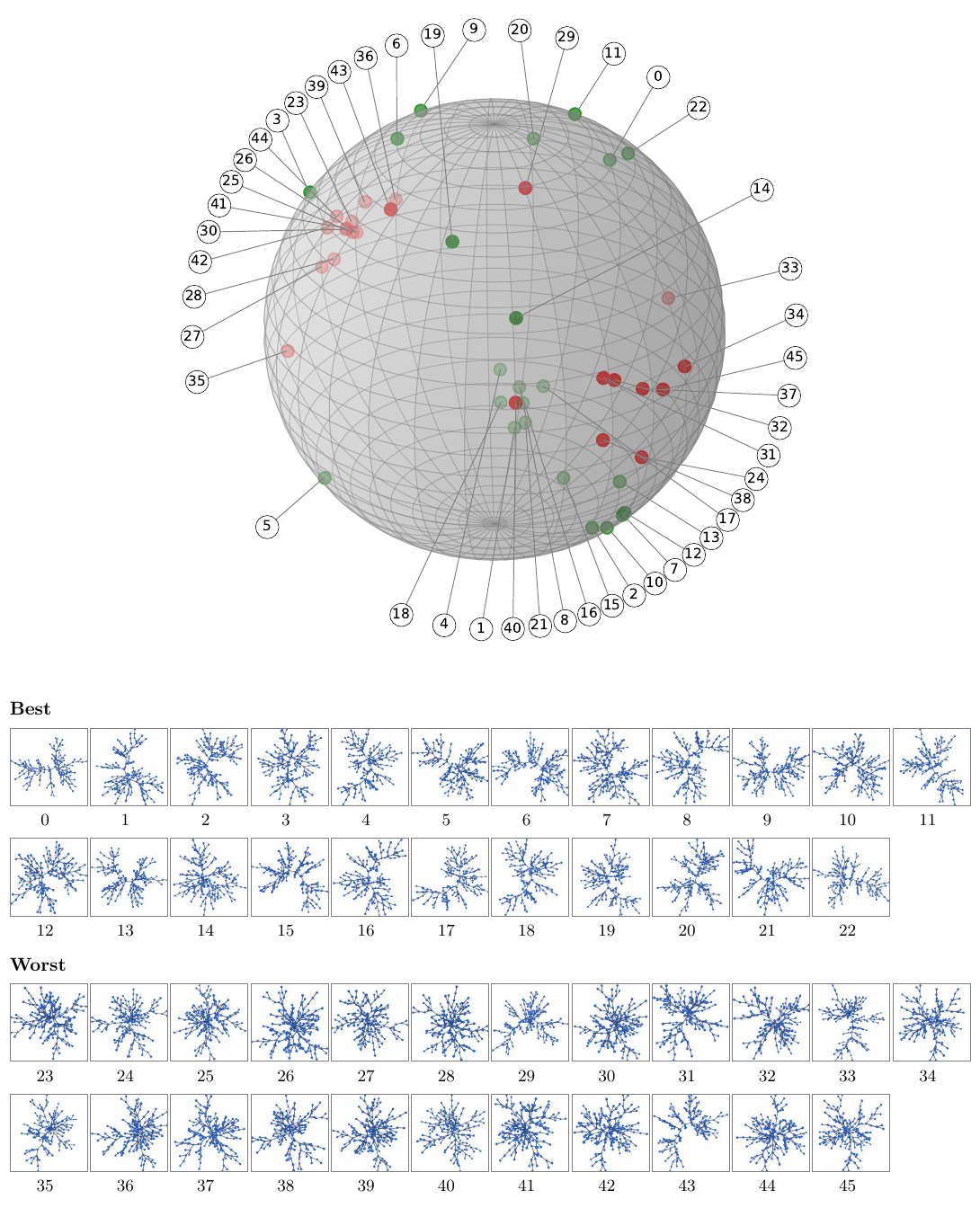}
    \caption{The distribution of all perspectives selected by users as \textit{best} (green) and \textit{worst} (red) mapped on a sphere surface for graph \textbf{E-10} (tree\_200\_2) with $|V| = 200, \; |E| = 199$.}
\end{figure}

\begin{figure}
    \centering
    \includegraphics[width=1\linewidth]{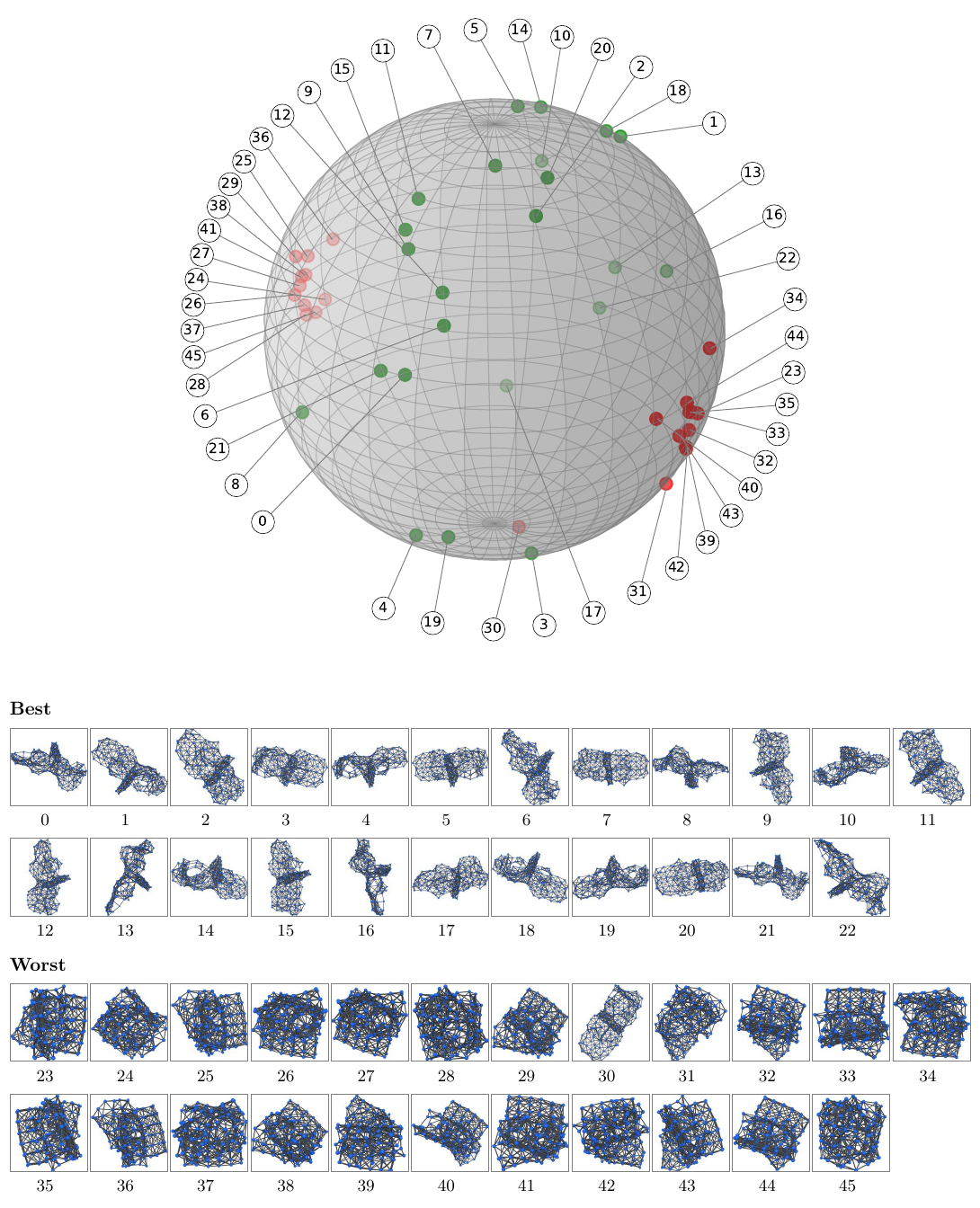}
    \caption{The distribution of all perspectives selected by users as \textit{best} (green) and \textit{worst} (red) mapped on a sphere surface for graph \textbf{E-11} (dwt\_209) with $|V| = 209, \; |E| = 976$.}
\end{figure}